\documentclass[preprint]{aastex63}
\usepackage{amsmath}
\usepackage{float}
\usepackage{booktabs, multirow} % for borders and merged ranges
\begin{document}

\newcommand{\scell}[3][c]{%
  \begin{tabular}[#1]{@{}#2@{}}#3\end{tabular}}

\title{\textbf{Grain Size Effects on UV-MIR (0.2-14 $\mu$m) Spectra of Carbonaceous Chondrite Groups}}
\accepted{2023 August 20}
\submitjournal{The Planetary Science Journal}

\author[0000-0001-6018-1729]{David Cantillo}
\affiliation{Lunar \& Planetary Laboratory, The University of Arizona, Tucson, AZ, USA}

\author[0000-0002-7743-3491]{Vishnu Reddy}
\affiliation{Lunar \& Planetary Laboratory, The University of Arizona, Tucson, AZ, USA}

\author[0000-0002-4412-5732]{Adam Battle}
\affiliation{Lunar \& Planetary Laboratory, The University of Arizona, Tucson, AZ, USA}

\author[0000-0003-1383-1578]{Benjamin N. L. Sharkey}
\affiliation{Lunar \& Planetary Laboratory, The University of Arizona, Tucson, AZ, USA}

\author[0000-0002-0183-1581]{Neil C. Pearson}
\affiliation{Planetary Science Institute, Tucson, AZ, USA}

\author[0000-0001-9418-1663]{Tanner Campbell}
\affiliation{Department of Aerospace \& Mechanical Engineering, The University of Arizona, Tucson, AZ, USA}
\affiliation{Lunar \& Planetary Laboratory, The University of Arizona, Tucson, AZ, USA}

\author[0000-0001-5766-8819]{Akash Satpathy}
\affiliation{Lunar \& Planetary Laboratory, The University of Arizona, Tucson, AZ, USA}

\author[0000-0003-2285-3074]{Mario De Florio}
\affiliation{Department of Systems \& Industrial Engineering, The University of Arizona, Tucson, AZ, USA}

\author[0000-0001-6076-8992]{Roberto Furfaro}
\affiliation{Department of Systems \& Industrial Engineering, The University of Arizona, Tucson, AZ, USA}
\affiliation{Department of Aerospace \& Mechanical Engineering, The University of Arizona, Tucson, AZ, USA}

\author[0000-0002-0764-4672]{Juan Sanchez}
\affiliation{Planetary Science Institute, Tucson, AZ, USA}

\correspondingauthor{David C. Cantillo}
\email{davidcantillo@arizona.edu}

\begin{abstract}
Carbonaceous chondrites are among the most important meteorite types and have played a vital role in deciphering the origin and evolution of our solar system. They have been linked to low-albedo C-type asteroids, but due to subdued absorption bands, definitive asteroid-meteorite linkages remain elusive. A majority of these existing linkages rely on fine-grained (typically $<$45 $\mu$m) powders across a limited wavelength range in the visible to near-infrared (VNIR, 0.35-2.5 $\mu$m). While this is useful in interpreting the fine-grained regolith of larger main belt objects like Ceres, recent spacecraft missions to smaller near-Earth asteroids (NEAs), such as Bennu and Ryugu, have shown that their surfaces are dominated by larger grain size material. To better interpret the surfaces of these smaller, carbonaceous NEAs, we obtained laboratory reflectance spectra of seven carbonaceous chondrite meteorite groups (CI, CM, CO, CV, CR, CK, C2-ungrouped) over the ultraviolet to mid-infrared range (0.2-14 $\mu$m). Each meteorite contained five grain size bins (45-1000 $\mu$m) to help constrain spectral grain size effects. We find a correlation between grain size and absolute reflectance, spectral slope, band depth, and the Christiansen feature band center. Principal component analysis of grain size variation illustrates a similar trend to lunar-style space weathering. We also show that the Bus-DeMeo asteroid taxonomic classification of our samples is affected by grain size, specifically shifting CM2 Aguas Zarcas from a Ch-type to B-type with increasing grain size. This has implications for the parent body of the OSIRIS-REx target, Bennu. With Aguas Zarcas, we present results from Hapke modeling.
\end{abstract}

\section{Introduction}
Meteorites provide a powerful tool in helping us understand the composition and early evolution of the solar system. When meteorites are linked to their parent bodies, a wealth of laboratory data can be derived and applied to the object that would not be possible without a sample return mission. Visible to near-infrared (VNIR) reflectance spectroscopy (0.35-2.5 $\mu$m) has been crucial in characterizing the surface composition of asteroids by deriving their surface mineralogy and enabling the identification of meteorite analogs. The first connections between meteorites and asteroids began in the 1970s with researchers noting similar albedos and weak absorption bands between carbonaceous chondrite meteorites and low albedo asteroids \citep{johnson1973optical}. More detailed spectral classification of asteroids \citep{tholen1984asteroid,bus2002phase, demeo2009extension} soon allowed for more robust comparisons with meteorite groups. Most spectral databases, such as RELAB \citep{pieters1983strength,pieters2004relab}, utilize fine-grained powders ($<$45 $\mu$m) that provide successful matches to larger bodies with a similar, fine-grained regolith such as Ceres \citep{nathues2016fc}. While these meteorite powders are adequate in matching the surface of larger planetary bodies, including large main belt objects, recent spacecraft observations of smaller NEAs have forced researchers to reconsider their assumptions regarding asteroid regolith and grain size \citep{watanabe2019hayabusa2,lauretta2019unexpected}.  \\

Near-Earth asteroids (NEAs) are unique in that they allow ground-based telescopes to study the physical characteristics of smaller asteroids ($<$10 km) that would normally be too faint to observe at main belt distances (2-3.5 AU). Due to the size and shape of their orbits, the NEA population is thought to be derived from well-established sources in the main belt that have been gravitationally perturbed through orbital resonances \citep{bottke2002debiased,granvik2018debiased}. There are additional size-dependent forces at play, with very small asteroids ($<$100 m) being transported more effectively into Earth-crossing orbits through non-gravitational processes \citep{bottke2006yarkovsky}. The proximity of these NEAs with the Earth requires accurate classification to assess impact risk, better understand the economic potential \citep{Sanchez2021metal} and drive scientific questions about their origin. Like with larger planetary bodies, this classification is largely done by comparing laboratory spectra of powders ($<$45 $\mu$m) to the telescopic spectrum of the object. Recent evidence, however, suggests that the surfaces of smaller asteroids may not be as fine-grained as originally thought \citep{yano2006touchdown,miyamoto2007regolith,lauretta2019unexpected,watanabe2019hayabusa2}. First, photometric observations of small NEAs often show rapid rotation, suggesting these bodies may be monolithic depending on their diameter \citep{whiteley2002lightcurve}. With this, the spectra of small asteroids contain relatively subdued absorption bands that match more closely with meteorite slabs rather than powders \citep{cloutis2013spectral}. Recent theoretical modeling conducted by \citet{maurel2016numerical} may explain how fine-grained regolith may be transported off the surface of small asteroids. The discrepancy in expected grain size on smaller asteroids was further widened with the recent spacecraft missions to small NEAs Ryugu (Hayabusa2) and Bennu (OSIRIS-REx). First observations of asteroid Ryugu showed a surface dominated by boulders and cobbles, devoid of sand- to pebble-sized regolith \citep{watanabe2019hayabusa2,grott2019low,sugita2019geomorphology}. Direct imaging of asteroid Bennu told a similar story: there is a high abundance of boulders on the surface and no large areas of well-sorted particulate regolith \citep{lauretta2019unexpected}. These results seem inconsistent with the moderate thermal inertia of the asteroid, possibly requiring fine layers of dust on the surface of the boulders \citep{dellagiustina2019properties}. \\

The unexpected, large grain sizes on the surfaces of these smaller, carbonaceous asteroids have forced researchers to revisit their original meteorite linkage methods that utilized fine-grained powder. Previous studies have shown that grain size can affect the spectral parameters of meteorites, including spectral slope, band depth, and albedo \citep{reddy2015mineralogy}, though there have been few systematic studies seeking to constrain the effect of grain size on an object’s spectrum. Recently, \citet{bowen2023grain} undertook a grain size study on ordinary chondrites and HED meteorites, which comprise about 90\% of all meteorites that fall on the Earth. Here, we examine the effects of grain size on the reflectance spectra of carbonaceous chondrites to expand on \citet{bowen2023grain} with the study of additional meteorite classes. Unlike HEDs and ordinary chondrites, which have well-defined absorption features in the VNIR, carbonaceous chondrites lack these diagnostic spectral features and are more difficult to link to their parent bodies. We expect that our grain size study of carbonaceous chondrites will aid in the classification of C-type NEAs that make up $\sim$20\% of the near-Earth object population \citep{perna2013near}. Our results may also aid in the interpretation of Bennu and Ryugu’s rougher surfaces, which are frequently linked to carbonaceous chondrite meteorites \citep{clark2011asteroid,hamilton2019evidence,yada2022preliminary,greenwood2022oxygen}. \\

In our experiment, we collected new UV-mid IR (0.2-14 $\mu$m) spectra of seven carbonaceous chondrite meteorites (CI, C2-UNG, CM, CO, CV, CR, CK, see Table \ref{table:sample_composition} for detailed list) across five different grain sizes. While many studies have looked at carbonaceous chondrites, few have specifically studied the effects of grain size, especially over a large wavelength range and with multiple meteorite groups. In a series of eight papers, Cloutis et al. conducted VNIR reflectance spectra of multiple meteorites in each of the main carbonaceous chondrite groups (e.g., \citet{cloutis2011spectral}, \citet{cloutis2011CM}). These carbonaceous chondrite surveys provided insight into the meteorite mineralogies and studied broad changes in physical characteristics including packing density and grain size. \citet{kiddell2018spectral} conducted a more comprehensive study of the grain size effects of carbonaceous chondrite powders on slabs covered with various powders, though this was only conducted in the VNIR wavelength range and did not include CI chondrites and ungrouped C2 chondrites. Extending this grain size study into a broader wavelength range (0.2-14 $\mu$m) may assist the interpretation of new asteroid spectra from space-based telescopes, including the recent commissioning of the James Webb Space Telescope (0.6-28.5 $\mu$m). To support this new wavelength coverage, there has been a significant interest in laboratory mid-infrared (MIR, 2.5-25 $\mu$m) spectra of carbonaceous chondrites, which includes multiple features and a wealth of mineralogical information. The phyllosilicate OH/H$_2$O feature at $\sim$3.0 $\mu$m has been linked to hydration in multiple studies (e.g. \citet{beck2014transmission,king2015characterising,takir2015toward,garenne2016bidirectional,takir2019hydration,potin2020style}), but its hydrated nature can make it more difficult to study in the laboratory and with ground-based telescopes due to atmospheric water. \citet{hanna2020distinguishing} found a shift in the $\sim$9 $\mu$m Christiansen feature in slab spectra due to aqueous alteration, and multiple researchers have noted changes in the shape and position of the $\sim$11.3 $\mu$m Si-O stretching band depending on the amount of aqueous alteration \citep{beck2014transmission,mcadam2015aqueous,hanna2020distinguishing, bates2020linking}. Most of these studies have utilized fine-grained powders, which can introduce transparency features that are absent in coarser powders or slabs \citep{shirley2019particle,hanna2020distinguishing} and are not as relevant to near-Earth asteroids. As a result, the understanding of grain size effects on carbonaceous chondrite spectra in this MIR region is crucial in determining how broadly these trends can be applied to planetary objects.

\section{Sample descriptions}

\subsection{Sample Selection}
Due to the difficulty of crushing metal, we chose to exclude the carbonaceous chondrite types that were high in metal content: CBs and CHs. This left us with six main carbonaceous chondrite groups (CI, CM, CO, CV, CR, and CK) as well as one ungrouped type (C2) that was made available to us. Samples were selected from our collection at the University of Arizona. 

\begin{table}[hb!]
    \centering
    \caption{Meteorite Composition}
    \label{tb1}
    \begin{tabular}{lccc}
        \textbf{Sample Name} & \textbf{Type} & \textbf{Major Minerals} &
        \textbf{Reference} \\
        
        \noalign{\smallskip}\hline\hline
        
        Orgueil & CI1 & \scell[c]{c}{Fe-Mg Serpentines, Saponite, Magnetite,\\                                          Carbonates, Sulfides} & 1 \\
        
        \hline\noalign{\smallskip}

        Tarda & C2-UNG & \scell[c]{c}{Smectite, Serpentine, \\
                                        Olivine, Diopside} & 2,3 \\

        \hline\noalign{\smallskip}

        Aguas Zarcas & CM2 & \scell[c]{c}{Cronstedite, Chrysotile, \\
                                          Ferrotochilinite, Calcite,
                                          Magnetite} & 4,5 \\

        \hline\noalign{\smallskip}

        Kainsaz & CO3.2 & \scell[c]{c}{Olivine, Enstatite, Pyroxenes, \\
                                        Anorthite, Spinel} & 2,4 \\

        \hline\noalign{\smallskip}

        Allende & CV3 & \scell[c]{c}{Olivine, Enstatite, High Ca-Pyroxene, \\
                                      Fassaite, Magnetite} & 1,4 \\

        \hline\noalign{\smallskip}

        NWA 10324 & CR2 & \scell[c]{c}{Olivine, low Ca-Pyroxene\\
                                            Kamacite, Troilite, Feldspathic Glass} & 1,2 \\

        \hline\noalign{\smallskip}

        NWA 10563 & CK4 & \scell[c]{c}{Olivine, low Ca-Pyroxene, \\
                                        Diopside, Fassasite, Anorthite} & 1,2 \\

        \noalign{\smallskip}\hline
    \end{tabular}
    \tablecomments{Summary of the meteorite samples used in this
              research and their types. Major minerals are representative of
              the type of meteorite and not necessarily specific to the sample. \\
              \textbf{References.} (1) \cite{rubin2021meteorite}; (2) Meteoritical Bulletin; (3) \cite{chennaoui2021}; (4) \cite{Brearley1998}; (5) \cite{Garvie2021}}
    % \caption{\textbf{References.} (1) \cite{rubin2021meteorite}; (2) Meteoritical Bulletin; (3) \cite{chennaoui2021}; (4) \cite{Brearley1998}; (5) \cite{Garvie2021} }
    \label{table:sample_composition}
\end{table}

\subsection{Orgueil (CI1)}

CI chondrites are distinct in that their non-volatile elemental composition is nearly identical to that of the Sun, the closest match of any meteorite group \citep{Anders1982solar}. While considered the most primitive material available for study, CI chondrites show evidence of aqueous alteration with hydrous sulfate veins \citep{dufresne1962chemical}. \citet{Clayton1984isotope} also cite unique oxygen isotopic composition as evidence of alteration of original anhydrous silicate material in a warm, wet environment. Unlike other carbonaceous chondrites, which contain various amounts of chondrules, CI chondrites are entirely composed of matrix material. \\

Orgueil is one of nine cataloged CI1 chondrite meteorites and one of only five observed as a fall, with approximately 14 kilograms of material recovered following its entry over southern France on May 14th, 1864 (Meteoritical Bulletin). With a petrologic type of 1, Orgueil has experienced a high degree of aqueous alteration resulting in very few chondrules. Instead, Orgueil is almost completely composed of matrix material, including Fe-bearing and Mg-rich serpentine and smectite \citep{tomeoka1988matrix} which compose approximately 60-65 wt\% of Orgueil \citep{kerridge1976major}. Magnetite (11.0 $\pm$ 0.4 wt\%) and carbon (4.88 wt\%) are also present at small grain sizes \citep{hyman1983magnetite,pearson2006carbon}. An additional major matrix material is ferrihydrite, a fine-grained material rich in Fe$^{3+}$ and coated with S and Ni. Petrographic studies of the olivine grains in the matrix reveal a generally low-Fe heterogeneous composition, peaking at Fa$_{1}$ though ranging up to Fa$_{55}$ \citep{mcsween1977nature}.

\subsection{Tarda (C2-UNG)}

Tarda is one of 27 C2-ungrouped meteorites discovered; and is one of only 52 observed falls across all carbonaceous chondrites, making it a particularly rare specimen (Meteoritical Bulletin). The fall was observed on August 25, 2020, over the Moroccan Sahara and the first pieces were collected the following day \citep{chennaoui2021, Marrocchi2021}. Tarda has a friable consistency, and many pieces were found to have broken apart when they landed on the ground. \\

As a petrologic type 2 meteorite, Tarda’s history includes a moderate amount of aqueous alteration. Evidence of this is seen in Tarda’s fine-grained matrix, dominated by phyllosilicates (large quantities of smectite and serpentine) with magnetite, Fe-sulfides, dolomite, and olivine \citep{Marrocchi2021, Applin2022}. Tarda’s total water content was found to be 8.54 wt\% \citep{Marrocchi2021}. Oxygen isotope measurements of Tarda all fall between those of CI and CY ("Yamato-type" group proposed by \citet{ikeda1992overview}) chondrites and are similar to the Tagish Lake meteorite \citep{chennaoui2021}. All observed olivine grains were dominated by forsterite (Fo) with geochemistry ranging from Fo$_{98.7}$ to Fo$_{99.3}$. Trace amounts of Ni-bearing pyrrhotite, chromite, and kamacite are also measurable in Tarda \citep{Marrocchi2021}.

\subsection{Aguas Zarcas (CM2)}
CM chondrites are an abundant group of carbonaceous chondrites with 725 unique meteorites (Meteoritical Bulletin). Petrologically, most fall as type 2, indicating a moderate amount of aqueous alteration but less than CI1 chondrites. While uncommon, there are CM1 members that contain more serpentine-group phyllosilicates and fewer calcium-aluminum inclusions \citep{cloutis2011CM}. Compared to CI chondrites, CM chondrites have a higher chondrule abundance ($\sim$20 vol\%) and a similar olivine composition \citep{mcsween1977nature}, though are still dominated by a phyllosilicate matrix \citep{Brearley1998}. Mineralogically, CM chondrites are shown to contain $\sim$9 wt\% of water bound in these phyllosilicates \citep{rubin2007progressive} as well as sparse calcium-aluminum inclusions (CAIs) and amoeboid olivine inclusions (AOIs) \citep{suttle_21}. Olivine and pyroxene are also abundant, with concentrations varying between 5.9 and 33.8 vol\% respectively. These anhydrous silicates are also present within the chondrules and are typically Fe-poor with values of Fa$_{<5}$ \citep{howard_09, howard_11, howard_15, Brearley1998}. \\

For our study, we selected CM2 Aguas Zarcas that fell on April 23, 2019, in Alajuela, Costa Rica. The fusion-crusted and blocky sample showed no apparent terrestrial aqueous alteration and was acquired prior to rainfall. The bulk composition of Aguas Zarcas is dominated by phyllosilicates, olivine and pyroxene grains, and Fe-Ni sulfides \citep{hicks2020aqueous, takir2020spectroscopy}. While officially classified as a CM2, Aguas Zarcas also contains at least four additional lithologies \citep{kerraouch2021polymict,kerraouch2022heterogeneous}. These varying lithologies, including metal-rich and C1-like, suggest different hydration periods and alteration throughout their evolution. A comprehensive spectral analysis of Aguas Zarcas provides a unique opportunity to study a CM2 sample with limited terrestrial weathering and is especially crucial given that CM chondrites are considered spectral analogs to Bennu, an NEA visited by the OSIRIS-REx spacecraft \citep{hamilton2019evidence}.

\subsection{Kainsaz (CO3.2)}
CO chondrites are also relatively abundant among carbonaceous chondrites, with approximately 720 meteorites cataloged in the Meteorical Bulletin. CO chondrites typically have a petrologic type of 3, with little to no phyllosilicates formed from aqueous alteration (\citep{RubinCO,Zolensky1993}. Instead, the matrix of CO chondrites consists largely of fine-grained olivine as well as additional pyroxene. Chondrules make up about half of CO3 chondrites, typically with low-Fe values between Fa$_{0}$\% to Fa$_{10}$\% though ranging up to Fa$_{48}$ \citep{Brearley1998,mcsween1977nature}. Other major constituents include AOIs, CAIs, Fe-Ni metal, and lithic fragments that are typically isolated olivine crystals (\citep{RubinCO,mcsween1977nature,McSweenCO}. As reviewed in \citet{CloutisCO}, several authors have further subdivided the CO chondrites based on metamorphic properties \citep[e.g.,][]{McSweenCO,ScottCO,SearsCO,ChizmadiaCO,GreenwoodCO}. Like main petrologic types, petrologic subtypes are numbered to correspond with increasing metamorphic processing, with known CO samples ranging from 3.1 - 3.8 (Meteoritical Bulletin Database). \\

For our study of CO chondrites, we obtained a sample of the meteorite Kainsaz, an observed fall on September 13, 1937, in Russia. Kainsaz, as a CO type 3.2, represents material towards the lower end of the metamorphic alteration sequence. Its matrix is heterogeneous, composed of fine grains of olivine, pyroxene, and FeNi metal \citep{KellerCO}. Compared with other CO chondrites, Kainsaz has a high total abundance of Fe-Ni metal at 14.6 wt\% \citep{RubinCO}. Olivine is present and is relatively Fe-rich with fayalite content ranging from Fa$_{48}$\% to Fa$_{72}$\%.

\subsection{Allende (CV3)}
There are approximately 660 known CV carbonaceous chondrite meteorites (Meteoritical Bulletin). CV chondrites, like COs, generally have a petrologic type of 3 that is a result of little to no aqueous alteration. Approximately 48 vol\% of CVs are made of chondrules that are largely Fe-poor but reaching up to Fa$_{48}$\% \citep{mcsween1977nature, Brearley1998}. The matrix of CVs, similar in abundance to its chondrules, is made up of mostly Fe-rich olivine (Fa$_{>48}$\%, \citep{kojima1996indicators,weisberg1998fayalitic,krot1998progressive}) grains as well as small amounts of pyroxene, pyrrhotite, and other minor constituents \citep{buseck1993matrices,Zolensky1993}. AOIs and CAIs vary in concentration among CVs, ranging from 1.2-8.6 vol\% and 0.3-9.4 vol\%, respectively \citep{grossman1976amoeboid}. Along with their petrologic type, CV chondrites can also be subdivided based on their oxidized (CV3$_{Ox}$) or reduced (CV3$_{R}$) state \citep{mcsween1977nature,krot1995mineralogical}. The oxidized group can be further subdivided based on the degree of aqueous alteration: Allende-like CV3$_{OxA}$ have little to no phyllosilicates, while Bali-like CV3$_{OxA}$ contain more evidence of hydration and less metal \citep{weisberg1998fayalitic}. \\

We selected CV3 Allende for our study, which was witnessed falling over the Chihuahua state, Mexico, on February 8, 1969 \citep{clarke1971allende}. Allende is largely composed of Fe-poor chondrules and an Fe-rich matrix that contains more than 80\% of fayalitic olivine and minor amounts of pyroxene \citep{weisberg1998fayalitic}. Allende is well-known for its CAI abundance and refractory materials which have helped understand the formation of the solar system and compose 2.2 vol\% of the meteorite \citep{kornacki1984petrography}.

\subsection{NWA 10324 (CR2)}
CR chondrites are somewhat uncommon, with only 206 known meteorites that have been cataloged (Meteorical Bulletin). CR chondrites are unique among carbonaceous chondrites in that they have a significantly reduced chemistry and contain a relatively high abundance (10 - 16 wt\%) of Fe-Ni metal \citep{cloutis2012CR} and magnetite \citep{kallemeyn1994compositional}. Like other carbonaceous chondrites, CR chondrites are largely composed of mafic silicate material consisting of low-Fe (Fa$_{4-6}$) olivine and small amounts of low-Fe pyroxene \citep{Weisberg1993CR, Brearley1998, Krot2002CR}. Compared to other carbonaceous chondrite meteorites, CR chondrites contain a higher amount of chondrules (50 - 60 vol\%, \citep{Weisberg2006}). These magnesium-rich chondrules contain much of the Fe-Ni metal and magnetite \citep{Krot2002CR} associated with CR chondrites. The matrix, consisting mostly of olivine, accounts for 30 - 50 vol\% \citep{WeisbergPrinz2000, Zolensky1993}. The remaining portion of CR chondrites consists of a small amount (0.1 - 2.6 vol\%) of refractory inclusions \citep{Aleon2002}. Most CR chondrites fall in petrologic types 1 - 3, indicating a range of aqueous alteration that results in varying abundances of hydrated phyllosilicates. \\

For our CR chondrite sample, we chose NWA 10324, a CR2 chondrite that was found in Northwest Africa in 2015 \citep{Bouvier2017a}. With a petrologic type of 2, NWA 10324 has received a moderate amount of aqueous alteration that makes it representative of most CR chondrites. Compared to other CR chondrites, NWA 10324 contains a lower amount of low-Fe olivine (Fa$_{3.3\pm1.3}$) but is well within the range that is typical for CR chondrites (Fa$_{4-6}$) considering the associated uncertainty.

\subsection{NWA 10563 (CK4)}
There are 511 CK chondrites that have been recorded in the Meteoritical Bulletin, making them a relatively abundant group. CK chondrites are highly oxidized for carbonaceous chondrites, having almost no Fe-Ni metal, a high modal abundance of magnetite (1.2-8.1 vol\%), and FeO olivine (Fa$_{28-33}$) \citep{Cloutis2012, Kallemeyn1991, Geiger1995}. Olivine is in both the matrix and chondrules, dominating the meteorite composition. CK meteorites are possibly linked to CVs and COs based on their oxygen isotopes, elemental abundance patterns, as well as some qualitative similarities \citep{Kallemeyn1991}. In addition to equilibrated olivine, CK matrices are also dominated by plagioclase, low-Ca pyroxene (Fs$_{23-29}$), and fine-grained magnetite \citep{Kallemeyn1991}. This matrix composition is similar to that of CVs and COs, but more highly metamorphosed. \\

Our choice for a CK chondrite is NWA 10563, a meteorite of petrologic type 4 found in Northwest Africa in 2015 \citep{Bouvier2017b}. Being of petrologic type 4, it has undergone moderate thermal alteration for a CK (typically types 3-6). This meteorite contains both orthopyroxene (Fs$_{25.1-25.2}$Wo$_{0.6-2.6}$) and clinopyroxene (Fs$_{10.8-14.7}$Wo$_{46.1-47.0}$), as well as equilibrated olivine (Fa$_{29.4-29.7}$, FeO/MnO = 95-105). There are well-formed chondrules containing Cr-magnetite as well as a finer-grained matrix of Cr-magnetite and plagioclase \citep{Bouvier2017b}. 

\section{Methodology}
\label{methodology}

\subsection{Sample Preparation}
Samples from large slices were first dry-cut with a diamond-bladed band saw to obtain a smaller sample piece and preserve material for future study. Samples of Orgueil, Tarda, and Aguas Zarcas were not found in large slices and therefore required special preparation. Our Orgueil sample consisted of several fragments enclosed in a vial, with most of the mass making up two fragments. Attempts were made at removing a very thin layer of fusion crust on the smaller of the two main fragments, but the fragile nature of the sample prevented us from removing the fusion crust entirely. Our sample of Tarda included minor amounts of contamination from quartz sand, natural fibers, and other terrestrial sources when we received it. Along with these terrestrial contaminants, we also identified moderate amounts of fusion crust on the sample. The largest of these contaminants were removed under the microscope with tweezers. From here, we selected large (1-3 mm) diameter pieces of Tarda until a sufficient mass was achieved. Rather than being a prepared slab, Aguas Zarcas was a broken stone with a 0.5-1 mm layer of fusion crust on three sides. The fusion crust was removed using a hardened high-carbon steel fine machinist's file. \\

Before crushing, photographs were obtained of each meteorite's slab or largest piece. Each sample (excluding Orgueil, see below) was then hand-crushed with an alumina mortar and pestle and dry-sieved in a stack to five grain size classes: 45-90, 90-150, 150-300, 300-500, and 500-1000 $\mu$m. One aspect to consider when sieving is the effect of hyperfine ($<$50 $\mu$m) particles clinging to the surfaces of larger grains. Work by \citet{mustard1997effects} and \citet{cooper2002midinfrared} have shown that these smaller particles can affect the mid-infrared region of a spectrum as the particle size approaches the wavelength of light in Mie scattering. While wet-sieving can reduce this effect, we chose not to further contaminate our samples with either water or alcohol. To understand the effect of clinging fines in our samples, we investigated photos of each of our powders with a 1:1 macro imaging lens with a 4 $\mu$m/px resolution at 600 nm. Throughout our samples, we did not observe a significant amount of clinging fines on our grains, though it is possible that very small grains below our resolution threshold may be present. In this case, there may be subtle effects on our Christiansen and Si-O stretching band features. Following a round of crushing and sieving, each grain size class was weighed with a calibrated precision balance with a precision of $\pm$ 0.0001 g. If 0.25 g of material was not obtained for each meteorite's grain size, additional crushing and sieving took place with the larger-grained, more abundant meteorite powders (see Table \ref{table:sample_mass} for detailed sample masses). Sample powders were photographed and then transferred into sample cups and leveled with a glass slide. \\

After measuring the spectra of our meteorite powders, we decided to include slab data for thoroughness and comparison. For our existing slabs (Kainsaz, Allende, NWA 10324, NWA 10563), this process was simple: slab spectra were taken on a spot immediately adjacent to material that had previously been cut for powdered samples. Our Aguas Zarcas sample, however, was crushed entirely. As a result, we selected a different large piece of Aguas Zarcas as our slab. Due to the fragile nature and the condition in which we received it, there was no available slab data for Tarda. Both irregularly shaped slab samples (Orgueil and Aguas Zarcas) were positioned in a sample cup and oriented to have their flattest face facing upwards. By doing so, we hoped to minimize changes in brightness and phase angle. \\

Due to the limited starting mass of our Orgueil sample (0.31 g), a different sample preparation method was conducted. Following photographs and spectral measurements of the slab, Orgueil was slowly and incrementally crushed to the next smaller grain size bin (500-1000 $\mu$m). Once the entire mass had been crushed and constrained to this next smaller grain size bin, the sample was weighed, prepared, and analyzed. Once spectral data was collected, the sample was returned to the mortar and pestle for additional crushing. This process continued successively down each grain size bin until the entire initial mass of Orgueil was constrained to 45-90 $\mu$m and $<$45 $\mu$m. The lowest mass achieved from this process was 0.14 g in the 90-150 $\mu$m grain size bin. \\

\begin{table}[H]
    \centering
    \caption{Meteorite Sample Grain Size Masses}
    \label{tb2}
    \begin{tabular}{cc}
    \addtolength{\tabcolsep}{-10pt}
    \begin{tabular}{lcc}
        \textbf{Sample Name} & \scell[c]{c}{\textbf{Grain Size} \\ 
                                            \textbf{($\mu$m)}} & 
        \scell[c]{c}{\textbf{Weight} \\ \textbf{(g)}} \\

        \noalign{\smallskip}\hline\hline

        Orgueil & \scell[c]{c}{45-90 \\ 90-150 \\ 150-300 \\ 300-500 \\ 
                             500-1000} & 
        \scell[c]{c}{0.16 \\ 0.14 \\ 0.18 \\ 0.16 \\ 0.21} \\

        \hline\noalign{\smallskip}

        Tarda & \scell[c]{c}{45-90 \\ 90-150 \\ 150-300 \\ 300-500 \\ 
                             500-1000} & 
        \scell[c]{c}{0.31 \\ 0.28 \\ 0.43 \\ 0.42 \\ 0.43} \\

        \hline\noalign{\smallskip}

        Aguas Zarcas & \scell[c]{c}{45-90 \\ 90-150 \\ 150-300 \\ 300-500 \\ 
                                   500-1000} & 
        \scell[c]{c}{0.42 \\ 0.34 \\ 0.58 \\ 0.75 \\ 2.12} \\

        \hline\noalign{\smallskip}

        Kainsaz & \scell[c]{c}{45-90 \\ 90-150 \\ 150-300 \\ 300-500 \\ 
                               500-1000} & 
        \scell[c]{c}{0.38 \\ 0.39 \\ 0.65 \\ 0.88 \\ 2.70} \\

        \noalign{\smallskip}\hline
    \end{tabular}

    \begin{tabular}{lcc}
        \textbf{Sample Name} & \scell[c]{c}{\textbf{Grain Size} \\ 
                                            \textbf{($\mu$m)}} & 
        \scell[c]{c}{\textbf{Weight} \\ \textbf{(g)}} \\

        \noalign{\smallskip}\hline\hline

        Allende & \scell[c]{c}{45-90 \\ 90-150 \\ 150-300 \\ 300-500 \\ 
                               500-1000} & 
        \scell[c]{c}{0.53 \\ 0.51 \\ 0.60 \\ 1.36 \\ 0.75} \\

        \hline\noalign{\smallskip}

        NWA 10324 & \scell[c]{c}{45-90 \\ 90-150 \\ 150-300 \\ 300-500 \\ 
                                 500-1000} & 
        \scell[c]{c}{0.42 \\ 0.41 \\ 0.74 \\ 1.11 \\ 1.43} \\

        \hline\noalign{\smallskip}

        NWA 10563 & \scell[c]{c}{45-90 \\ 90-150 \\ 150-300 \\ 300-500 \\ 
                                 500-1000} & 
        \scell[c]{c}{0.49 \\ 0.49 \\ 0.91 \\ 1.30 \\ 2.43} \\

        \noalign{\smallskip}\hline
    \end{tabular} \\
    \end{tabular}

    % \captionsetup{width=0.85\textwidth, justification=justified, 
    %               singlelinecheck=false}
    \tablecomments{Summary of meteorite sample grain sizes and 
              their masses.}
    \label{table:sample_mass}
\end{table}

\subsection{Spectral Measurements and Spectral Calibration}
Three different spectrometers and four different light sources were used to cover the measured wavelength range of 0.2-14 $\mu$m. The 0.2-1.1 $\mu$m wavelength (UV-Vis) region was measured by an Avantes AvaSpec-ULS2048x64TEC-EVO spectrometer with a spectral resolution of 1.4 nm FWHM. This spectrometer used a combined deuterium arc lamp and halogen lamp to provide illumination for the entire spectral range, with the deuterium arc lamp emitting in the 0.2-0.5 $\mu$m range and the halogen lamp emitting in the 0.4-1.1 $\mu$m  range. A fiber optic probe was used to transmit light to the sample and collect the light from it. The orientation of the fibers was incidence of 0$^\circ $, emission of 0$^\circ $, and total phase angle of 0$^\circ $. The spectra were converted to reflectance by first taking a white reference relative to a Spectralon$^{TM}$ panel and dividing the sample spectra by the reference. The 0.35-2.5 $\mu$m region (Vis-NIR) was collected using an Analytical Spectral Devices LabSpec 4 with a spectral resolution of 3 nm from 0.35-1.0 $\mu$m and 6 nm from 1.0-2.5 $\mu$m. Light was collected and delivered to the spectrometer using a fiber optic bundle with a custom quartz tungsten halogen lamp as the light source. This spectral range also used a Spectralon white reference to convert measurements to reflectance. The light source was at an incidence angle of 0$^\circ $ and the spectrometer fiber was at an emission angle of 30$^\circ $, for a phase angle of 30$^\circ $. For the 1.5-14 $\mu$m (NIR-MIR) region, a Thermofisher Nicolet FTIR spectrometer was used with a Mercury Cadmium Telluride detector (MCT-D), potassium bromide beam splitter, and a silicon nitride globar light source. Light was directed onto and collected from the sample using a SpecAc diffuse reflectance goniometer with an incidence at 0$^\circ $ and emission at 30$^\circ $. For this spectral range, a piece of flat 6061 aluminum alloy was ground with 45 $\mu$m grit and then vacuum coated with gold was used as the reference. Spectralon no longer has a flat reflectance curve beyond 2.5 $\mu$m while gold does. All spectral data were taken under ambient conditions. \\

To remove Spectralon$^{TM}$ absorption artifacts in the 0.2-0.4 $\mu$m and 2.21 $\mu$m region, the sample spectra were multiplied by a known calibrated reflectance spectra of Spectralon$^{TM}$ (see \citet{Kokaly_2017} and \citet{clark1990high}).  This was not done for the 1.5-14 $\mu$m sample spectra because gold has no sharp absorption features. The 0.35-2.5 $\mu$m spectra were corrected for minor offsets at detector transitions at 1.0 $\mu$m and 1.8 $\mu$m using the 1-1.8 $\mu$m region as this has been shown to be the most stable detector \citep{salisbury1998spectral}. The UV-VIS and the NIR-MIR spectra were then scaled to the VIS-NIR spectra. \\

Once the spectral data were reduced, band parameter analysis could be conducted. To measure band centers, areas, and depths, we utilized a python script described by \citet{Sanchez2017}. For each suspected absorption feature, a point on each side of the band was selected to derive a linear continuum. From here, two additional points around the reflective minimum were chosen to constrain a third-order polynomial fit. Band depths greater than a 0.5\% threshold below the linear continuum qualified as an absorption feature and were recorded. While the uncertainty in our laboratory reflectance is very small, error can be introduced in the selection of the continuum points. To account for this, we ran this band parameter program five times for every sample measurement using slightly varying continuum points. The mean values of these band parameters for our 45-90 $\mu$m samples are presented in Tables \ref{table:band_parameters_1} and \ref{table:band_parameters_2}, with the uncertainty being reported as two standard deviations of our five measured values. A modified version of this script was also utilized to select the absolute minimum and maximum values that were used to find the Christiansen Feature ($\sim$8.5-9.5 $\mu$m) and Si-O stretching band peak ($\sim$9.5-11.5 $\mu$m), respectively. In some cases, there were multiple Si-O stretching band peaks within the 9.5-11.5 $\mu$m region. To consistently track this feature, we found the reflectance maximum in the 45-90 $\mu$m sample between 9.5-11.5 $\mu$m and tracked it with increasing grain size, even though other nearby features had higher reflectance values in Orgueil, Tarda, and Kainsaz (see Figures \ref{fig:absolute_panels} and \ref{fig:relative_panels}). Since these are absolute values with very small instrument error, no uncertainty values are provided. Spectral slope measurements were conducted by taking the reflectance values at 1.8 and 0.8 $\mu$m, which are relatively featureless areas of meteorite spectrum that have been used by \citet{Cloutis2010} and \citet{Cantillo2021}.

\subsection{Compositional Variation}
The sieving procedure used for most of our meteorites (excluding Orgueil) was effective but may have certain limitations. Carbonaceous chondrites are composed of various components, ranging from fragile phyllosilicates to more resistant silicates and metal grains. The heterogeneous nature of these samples and their differing resistance to crushing may result in non-homogeneous material sorting throughout our grain size sieve stack. This possible uncertainty has been noted in other studies (e.g., \citet{cloutis2011CM}) and may result in minor spectral variability within our samples. When comparing spectral metrics between different meteorite samples (Tables \ref{table:band_parameters_1},\ref{table:band_parameters_2}), a single grain size group (45-90 $\mu$m) was utilized to minimize this possible variability.

\section{Results}
\begin{figure}[H]
    \centering
    \includegraphics[width=0.8\linewidth]{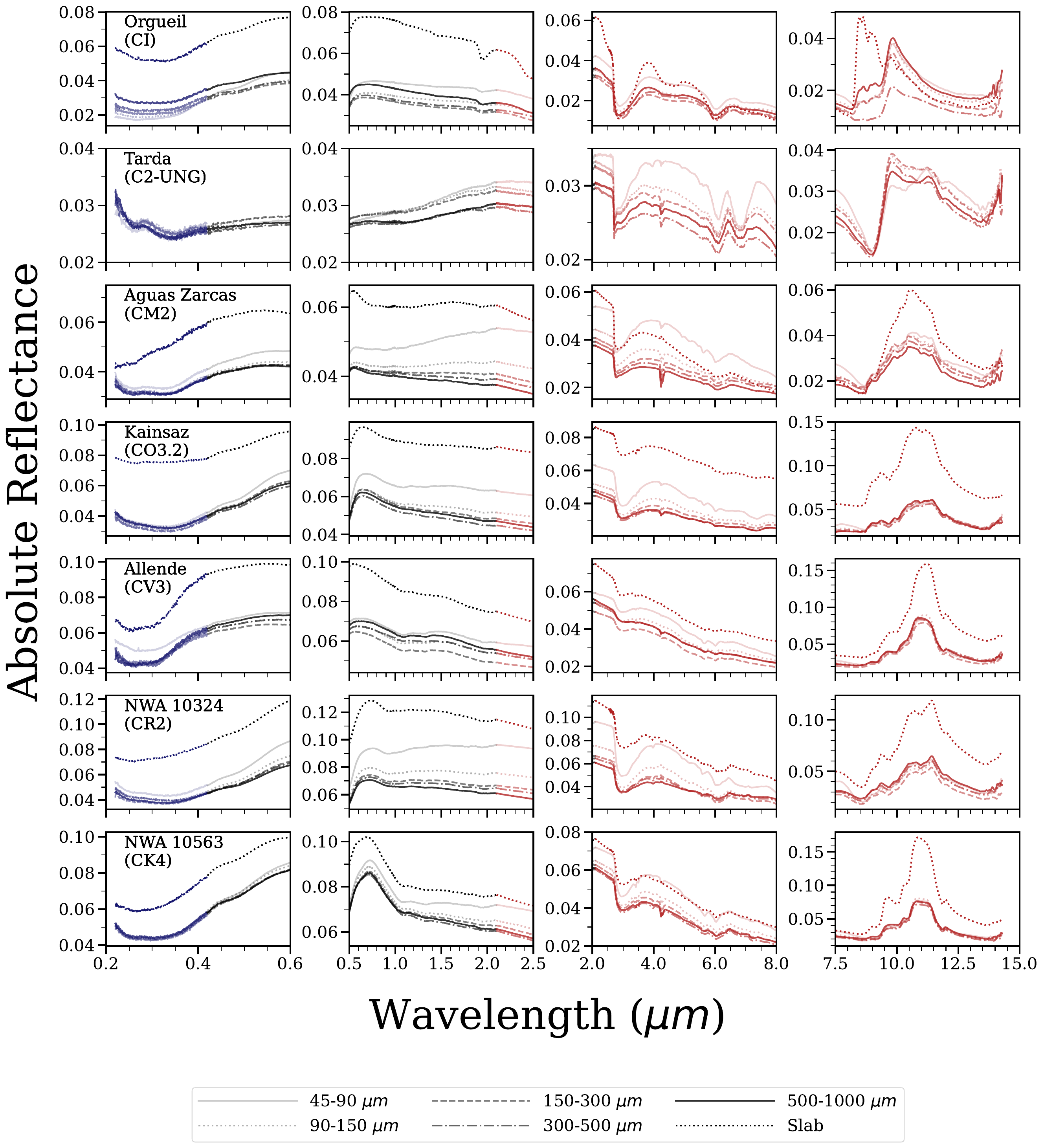}
    \caption{Absolute reflectance of the six grain size bins from our seven studied meteorites across the 0.2-14 $\mu$m wavelength range. Each row corresponds to an individual meteorite that is labeled in the leftmost column. Each column corresponds to a set wavelength range, consisting of 0.20-0.60 $\mu$m (Column 1), 0.5-2.5 $\mu$m (Column 2), 2.0-8.0 $\mu$m (Column 3), and 7.5-15.0 $\mu$m (Column 4) going from left to right. The range in reflectance for each subpanel varies across meteorite and wavelength range to best display the reflectance features. Different grain sizes across the same meteorite are noted by the line styling in the bottom legend. Color is utilized to indicate the spectrometer used for data collection: blue is Avantes UV, black is ASD VNIR, and red is FTIR NIR-MIR.}
    \label{fig:absolute_panels}
\end{figure}

\begin{figure}[H]
    \centering
    \includegraphics[width=0.8\linewidth]{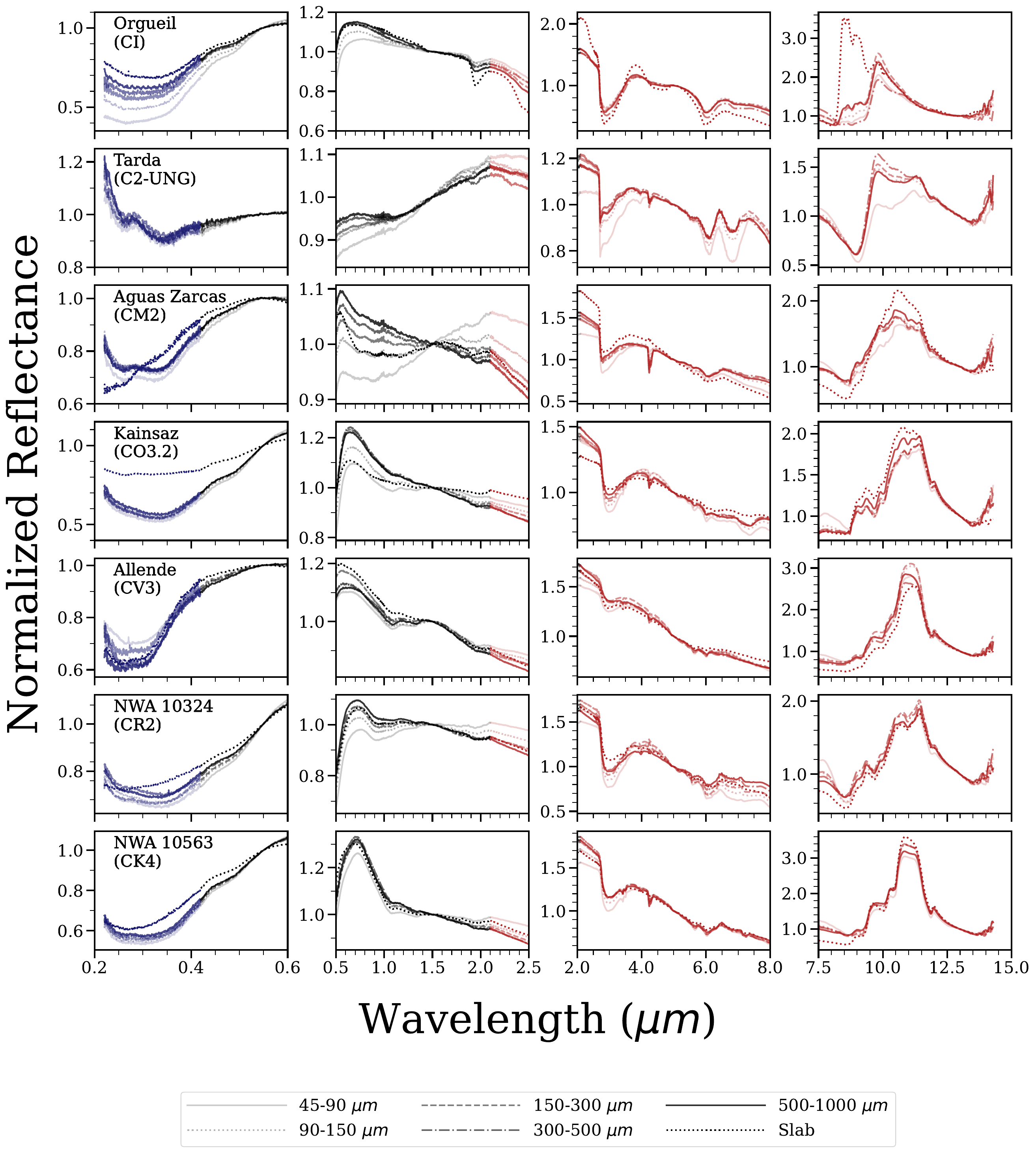}
    \caption{Relative reflectance of the six grain size bins from our seven studied meteorites across the 0.2-14 $\mu$m wavelength range. The figure layout is identical to that of Figure \ref{fig:absolute_panels}. Column 1 ($\sim$0.20-0.60 $\mu$m) is normalized to unity at 0.55 $\mu$m, Column 2 (0.6-2.5 $\mu$m) is normalized to unity at 1.5 $\mu$m, Column 3 (2.5-8 $\mu$m) is normalized to unity at 5.0 $\mu$m, and Column 4 ($\sim$8-14 $\mu$m) is normalized to unity at 13 $\mu$m. Normalizing the spectra in our four wavelength bins allows us to better see trends in slope and band depth across varying grain sizes.}
    \label{fig:relative_panels}
\end{figure}

The absolute and relative reflectance spectra of the five grain sizes for each of our six studied meteorites across the $\sim$0.2-14 $\mu$m range are shown in Figures \ref{fig:absolute_panels} and \ref{fig:relative_panels}. From here, more focused reflectance, spectral slope, and band feature analysis are performed to better understand the effects of grain size and compare the different meteorites.

\subsection{Spectral Band Features Across Meteorite Types}
The spectral data for our meteorites and their grain sizes cover a large wavelength range ($\sim$0.2-14 $\mu$m) containing numerous absorption and emission features. Due to the sheer volume of data available, it is best to incrementally work our way up from shorter to longer wavelengths and initially focus on the absolute reflectance data in Figure \ref{fig:absolute_panels} to characterize the band features. All band depth percentages presented in this subsection come from the 45-90 $\mu$m grain size bin of each meteorite (shown in Tables \ref{table:band_parameters_1} and \ref{table:band_parameters_2}) to focus more on broad meteorite comparisons. Analysis of the variation across grain size will be discussed later in the Results section. \\

Column 1 (0.20-0.60 $\mu$m) of Figure \ref{fig:absolute_panels} contains three separate absorption features seen across the meteorite spectra. First, all studied meteorites contain a shallow 0.25 $\mu$m carbon absorption feature that is most defined with ungrouped C2 Tarda (band depth 7.7\% $\pm$ 0.5\%) and CM2 Aguas Zarcas (band depth 8.2\% $\pm$ 1.4\%, Table 3). At 0.35 $\mu$m, a broad absorption feature is present in all samples that is likely due to weathering. CO3.2 Kainsaz has the deepest absorption at 16.5\% $\pm$ 1.3\%, with Tarda and Aguas Zarcas having the shallowest absorptions of 8.0\% $\pm$ 0.2\% and 9.0\% $\pm$ 1.1\%, respectively. In the visible range, a moderate iron oxide weathering feature at 0.50 $\mu$m is present in all meteorites but most prominent in Kainsaz (6.2\% $\pm$ 0.6\%), CR2 NWA 10324 (5.6\% $\pm$ 0.9\%), and CK4 NWA 10563 (3.9\% $\pm$ 0.4\%). Absolute reflectance also varies significantly among the meteorite groups. NWA 10324 has the brightest absolute reflectance at 0.55 $\mu$m ($\sim$8\%), while Tarda is significantly darker with the lowest reflectance of $<$3\%. \\

Column 2 (0.5-2.5 $\mu$m) of Figure \ref{fig:absolute_panels} contains two main absorption features at $\sim$1.0 and 1.9 $\mu$m. First, the 1.0 $\mu$m absorption feature is a result of Fe$^{2+}$ cations in silicates/phyllosilicates, commonly in the form of pyroxene, olivine, and/or magnetite. NWA 10563 has the deepest 1.0 $\mu$m absorption (11.4\% $\pm$ 0.1\%), while Tarda (2.6\% $\pm$ 0.3\%), and Aguas Zarcas (2.1\% $\pm$ 0.4\%, Table 3) have the weakest. The 1.9 $\mu$m band is indicative of pyroxene and is the deepest in CI1 Orgueil (3.7\% $\pm$ 0.2\%). In addition to these two prominent absorption features, there are an additional three minor absorption features in the VNIR range. Aguas Zarcas is the only meteorite to exhibit a 0.70 $\mu$m absorption that is indicative of hydrated phyllosilicates. With this, Orgueil displays a subtle absorption at 1.45 $\mu$m due to its high hydration. Lastly, there is a 2.3 $\mu$m Mg-OH/carbonate feature present in our more oxidized meteorites, Orgueil, Tarda, and Aguas Zarcas.  \\

Column 3 (2.0-8.0 $\mu$m) of Figure \ref{fig:absolute_panels} contains two additional absorption features observed across all meteorites. The 2.8 $\mu$m band is the result of hydrated phyllosilicates and is seen across all samples and grain sizes at moderate depths. This feature (as well as 6.0 $\mu$m) is strongly affected by absorbed water from the atmosphere. Orgueil has the deepest 2.8 $\mu$m absorption of 59.6\% $\pm$ 0.4\%, while Allende is the weakest among the meteorites at a more moderate depth of 10.0\% $\pm$ 0.4\% (Table 3). Further into the infrared at 6.0 $\mu$m is an additional water feature that is also present among all our studied meteorites. Orgueil has a very strong 6.0 $\mu$m band depth of 41.3\% $\pm$ 0.7\% while Kainsaz has the weakest depth at 7.6\% $\pm$ 0.1\%. There are additional, weaker bands in this wavelength range that are only present in specific meteorites. Near 3.4 $\mu$m, an absorption feature is present in Orgueil, Tarda, and Aguas Zarcas that is representative of organics. All the meteorites, excluding Orgueil, contain a relatively weak water absorption at 5.65 $\mu$m that is most prominent in NWA 10563 (7.24\% $\pm$ 0.7\%). At 6.9 $\mu$m, an additional carbonate band is present across Orgueil, Tarda, NWA 10324, and NWA 10563, with the largest absorption of 18.2\% $\pm$ 0.5\% seen in NWA 10324. A subtle 7.15 $\mu$m organics feature is also observed in all meteorites excluding Kainsaz, with Aguas Zarcas having the strongest of 3.4\% $\pm$ 0.7\%. Lastly, Kainsaz is the only meteorite with a moderate absorption (10.6$\pm$ 0.4\%) feature at 7.35 $\mu$m due to pyroxene/diopside. \\

Column 4 (7.5-15.0 $\mu$m) of Figure \ref{fig:absolute_panels} contains mid-infrared information on mineralogy and petrology. In this range, there are two prominent features. The first is the Christiansen Feature (C.F.), a reflective minimum and emission maximum that is often used to gauge the silica content of a material \citep{logan1973compositional}. The shortest wave C.F. minimum is found in Aguas Zarcas and Kainsaz (8.711 $\mu$m), while NWA 10563 has the longest at 9.149 $\mu$m. The second feature is known as the Si-O stretching band (Si-O S.B.), a reflective maximum and emission minimum \citep{salisbury1992emissivity}. The shortest Si-O S.B. peak is found in Orgueil (9.816 $\mu$m), though more typical Si-O S.B. peaks are found near $\sim$11 $\mu$m, with NWA 10324 having the longest wavelength peak at 11.409 $\mu$m. In addition to the Si-O S.B. peak, Si-O S.B. contrast was measured as the negative band depth percentage of the feature. NWA 10563 has the largest Si-O S.B. contrast of 251.1\% $\pm$ 4.0\%, while Tarda has the smallest (48.4\% $\pm$ 4.9\%).

\begin{table}[H]\centering
\caption{UV-Mid IR Meteorite Band Parameters for 45-90 $\mu$m Samples}
\addtolength{\tabcolsep}{5pt}
\label{table:band_parameters_1}
\begin{tabular}{lrrrrr}\toprule
\textbf{} &\textbf{Attribution} &\textbf{Orgueil} &\textbf{Tarda} &\textbf{Aguas Zarcas} \\
\textbf{} & &(CI1) &(C2-UNG) &(CM2) \\
\noalign{\smallskip}\hline\hline
\textbf{0.25 Depth (\%)} &C &6.82 $\pm$ 1.47 &7.74 $\pm$ 0.54 &8.18 $\pm$ 1.38 \\
\textbf{0.35 Depth (\%)} &Fe-O &19.80 $\pm$ 0.93 &8.04 $\pm$ 0.18 &9.04 $\pm$ 1.12 \\
\textbf{0.5 Depth (\%)} &Magnetite &3.22 $\pm$ 0.09 &2.46 $\pm$ 0.81 &1.22 $\pm$ 0.95 \\
\textbf{0.7 Depth (\%)} &Phyllosilicates &N/A &N/A &0.68 $\pm$ 0.09 \\
\textbf{0.9-1.25 Depth (\%)} &Px/Ol/Mag &0.94 $\pm$ 0.27 &2.62 $\pm$ 0.30 &2.06 $\pm$ 0.36 \\
\textbf{0.9-1.25 Center ($\mu$m)} &Px/Ol/Mag &1.249 $\pm$ 0.01 &1.119 $\pm$ 0.01 &1.087 $\pm$ 0.01 \\
\textbf{1.45 Depth (\%)} &OH/H2O &0.70 $\pm$ 0.20 &N/A &N/A \\
\textbf{1.9 Depth (\%)} &OH/H2O &3.70 $\pm$ 0.24 &1.16 $\pm$ 0.23 &0.94 $\pm$ 0.11 \\
\textbf{2.3 Depth (\%)} &Mg-OH, CO3 &0.70 $\pm$ 0.37 &0.64 $\pm$ 0.23 &0.68 $\pm$ 0.30 \\
\textbf{2.8 Depth* (\%)} &OH/H2O &59.64 $\pm$ 0.39 &24.44 $\pm$ 0.87 &33.18 $\pm$ 0.36 \\
\textbf{3.3-3.5 Depth (\%)} &Organics &2.88 $\pm$ 0.46 &1.46 $\pm$ 0.33 &1.68 $\pm$ 0.50 \\
\textbf{5.65 Depth (\%)} &H2O &N/A &1.20 $\pm$ 0.40 &5.66 $\pm$ 0.46 \\
\textbf{6.0 Depth* (\%)} &H2O &41.32 $\pm$ 0.70 &13.84 $\pm$ 0.18 &13.18 $\pm$ 0.22 \\
\textbf{6.9 Depth (\%)} &CO3 &5.88 $\pm$ 0.68 &18.18 $\pm$ 0.50 &N/A \\
\textbf{7.15 Depth (\%)} &Organics &1.36 $\pm$ 0.69 &0.60 $\pm$ 0.45 &3.40 $\pm$ 0.68 \\
\textbf{7.35 Depth (\%)} &Px/Diopside &N/A &N/A &N/A \\
\textbf{C.F. Center ($\mu$m)} &Si-O &8.785 &9.042 &8.711 \\
\textbf{Si-O S.B. Center ($\mu$m)} &Si-O &9.816 &11.212 &10.642 \\
\textbf{Si-O S.B. Contrast (\%)} &Si-O &110.52 $\pm$ 5.36 &48.40 $\pm$ 4.93 &98.44 $\pm$ 3.93 \\
\bottomrule
\end{tabular}
\end{table}

\begin{table}[H]\centering
\caption{UV-Mid IR Meteorite Band Parameters for 45-90 $\mu$m Samples, Continued}
%\addtolength{\tabcolsep}{-5pt}
\label{table:band_parameters_2}
\begin{tabular}{lrrrrrr}\toprule
\textbf{} &\textbf{Attribution} &\textbf{Kainsaz} &\textbf{Allende} &\textbf{NWA 10324} &\textbf{NWA 10563} \\
\textbf{} & &(CO3.2) &(CV3) &(CR2) &(CK4) \\
\noalign{\smallskip}\hline\hline
\textbf{0.25 Depth (\%)} &C &7.06 $\pm$ 1.57 &5.22 $\pm$ 1.15 &7.10 $\pm$ 0.62 &6.68 $\pm$ 1.92 \\
\textbf{0.35 Depth (\%)} &Fe-O &16.50 $\pm$ 1.28 &9.88 $\pm$ 1.55 &14.3 $\pm$ 0.85 &12.46 $\pm$ 0.59 \\
\textbf{0.5 Depth (\%)} &Magnetite &6.24 $\pm$ 0.64 &1.30 $\pm$ 0.49 &5.60 $\pm$ 0.89 &3.88 $\pm$ 0.38 \\
\textbf{0.7 Depth (\%)} &Phyllosilicates &N/A &N/A &N/A &N/A \\
\textbf{0.9-1.25 Depth (\%)} & Px/Ol/Mag &5.44 $\pm$ 0.11 &6.48 $\pm$ 0.09 &4.56 $\pm$ 0.11 &11.42 $\pm$ 0.09 \\
\textbf{0.9-1.25 Center ($\mu$m)} & Px/Ol/Mag &1.050 $\pm$ 0.01 &1.073 $\pm$ 0.01 &0.935 $\pm$ 0.03 &1.058 $\pm$ 0.01 \\
\textbf{1.45 Depth (\%)} &OH/H2O &N/A &N/A &N/A &N/A \\
\textbf{1.9 Depth (\%)} &OH/H2O &1.66 $\pm$ 0.23 &2.72 $\pm$ 0.09 &1.38 $\pm$ 0.22 &1.72 $\pm$ 0.36 \\
\textbf{2.3 Depth (\%)} &Mg-OH, CO3 &N/A &N/A &N/A &N/A \\
\textbf{2.8 Depth* (\%)} &OH/H2O &33.62 $\pm$ 0.09 &9.98 $\pm$ 0.43 &43.64 $\pm$ 0.54 &29.52 $\pm$ 0.43 \\
\textbf{3.3-3.5 Depth (\%)} &Organics &4.28 $\pm$ 0.46 &1.38 $\pm$ 0.38 &2.44 $\pm$ 0.61 &5.98 $\pm$ 0.55 \\
\textbf{5.65 Depth (\%)} &H2O &4.68 $\pm$ 0.22 &6.06 $\pm$ 0.58 &6.68 $\pm$ 0.86 &7.24 $\pm$ 0.70 \\
\textbf{6.0 Depth* (\%)} &H2O &7.56 $\pm$ 0.90 &8.04 $\pm$ 0.41 &18.96 $\pm$ 0.30 &14.06 $\pm$ 0.81 \\
\textbf{6.9 Depth (\%)} &CO3 &N/A &N/A &6.32 $\pm$ 1.00 &2.96 $\pm$ 0.33 \\
\textbf{7.15 Depth (\%)} &Organics &N/A &2.18 $\pm$ 0.54 &3.12 $\pm$ 1.13 &3.20 $\pm$ 0.62 \\
\textbf{7.35 Depth (\%)} &Px/Diopside &10.62 $\pm$ 0.43 &N/A &N/A &N/A \\
\textbf{C.F. Center ($\mu$m)} &Si-O &8.711 &8.553 &8.578 &9.149 \\
\textbf{Si-O S.B.  Center ($\mu$m)} &Si-O &11.403 &11.104 &11.409 &10.871 \\
\textbf{Si-O S.B. Contrast (\%)} &Si-O &84.92 $\pm$ 2.36 &220.66 $\pm$ 3.77 &112.08 $\pm$ 5.54 &251.12 $\pm$ 3.95 \\
\bottomrule
\end{tabular}
% \captionsetup{width=0.75\textwidth, justification=justified, singlelinecheck=false}
\tablecomments{Sample reflectance feature measurements. 
Reported values are averages of 5 measurements with 2 standard 
deviation (95\%) uncertainty. Band depths less than 0.5\% are denoted by N/A. The attribution of spectral features was conducted using the USGS spectral library \citep{Kokaly_2017}. \\
\textbf{*} Affected by absorbed water under ambient laboratory conditions.}
\end{table}

\subsection{Reflectance Across Grain Size}
\begin{figure}[H]
    \centering
    \includegraphics[width=0.8\linewidth]{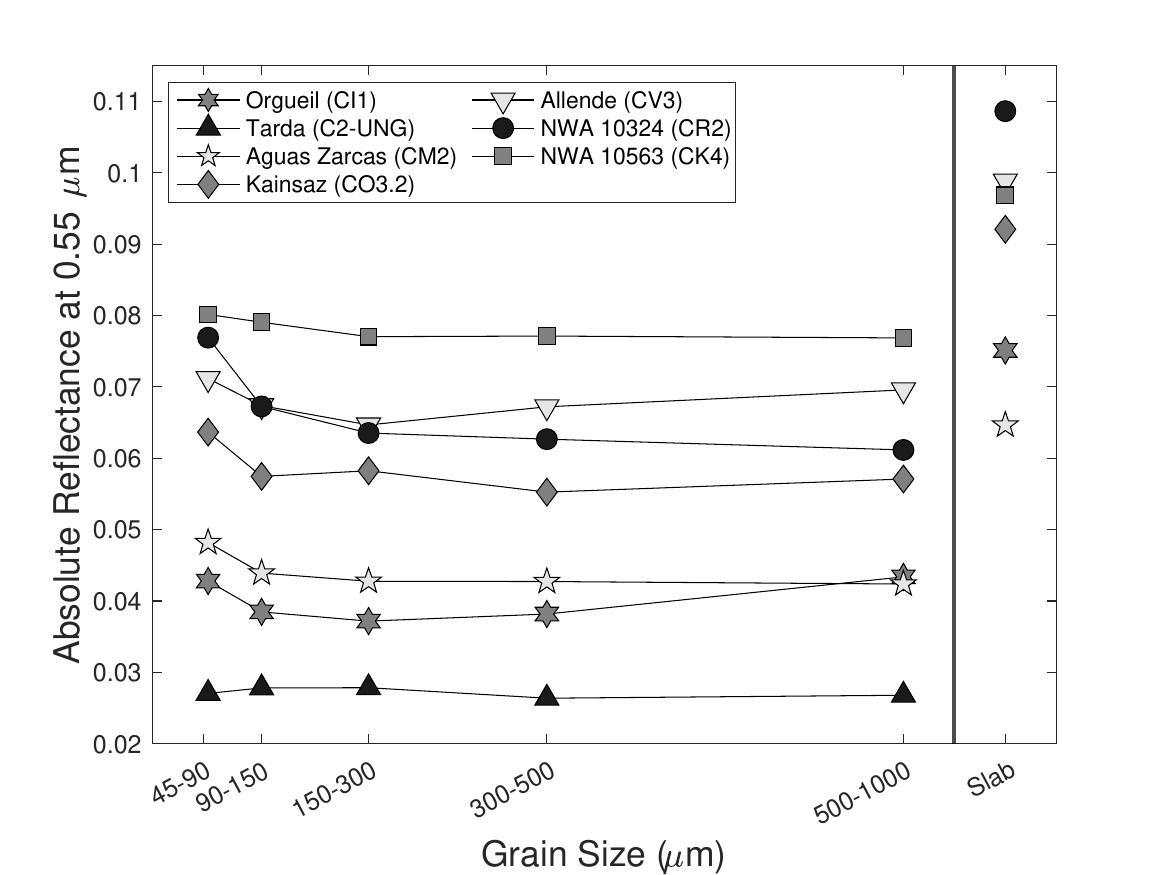}
    \caption{Plot of 0.55 $\mu$m (photometric V filter) absolute reflectance of our seven studied meteorite samples across our five grain size bins. Slab data are included for thoroughness, though are not connected due to their unconstrained grain size. Due to material constraints, Tarda lacks a slab measurement.}
    \label{fig:albedo_grainsize}
\end{figure}

Once band features have been identified among different meteorites, we can begin to compare the variation in grain size among these same meteorites starting with the simplest parameter: absolute reflectance. In Figure \ref{fig:absolute_panels}, we observe that most of the meteorites in Columns 1-3 ($\sim$0.20-8.0 $\mu$m) have a higher absolute reflectance with smaller grain sizes. This general relationship can be observed at 0.55 $\mu$m (photometric V filter) in Figure \ref{fig:albedo_grainsize}. At this wavelength, our samples generally exhibit a concave-up shape, with the highest reflectance at 45-90 $\mu$m with a subtle increase in reflectance at 500-1000 $\mu$m. All seven meteorites are darkest at moderate grain sizes between 150-1000 $\mu$m. Orgueil and Allende especially show a concave-up uptick in reflectance at larger grain sizes while Tarda remains relatively constant in reflectance across all grain size bins. Tarda has the lowest reflectance ($<$3\%) across all grain sizes while NWA 10563 is the brightest across constrained grain sizes at $\sim$8\%. All samples show significant upticks in their slab reflectance, though the unconstrained grain size of the slabs makes it difficult to interpret this trend. It is possible that this increase in slab reflectance may be due to more forward-scattering of light as a non-Lambertian reflector, unlike our more isotropic and diffuse powders. Past 8.0 $\mu$m (Column 4) in Figure \ref{fig:absolute_panels}, there is no clear trend in absolute reflectance with grain size. 

\subsection{Spectral Slope Across Grain Size}
\begin{figure}[H]
    \centering
    \includegraphics[width=0.8\linewidth]{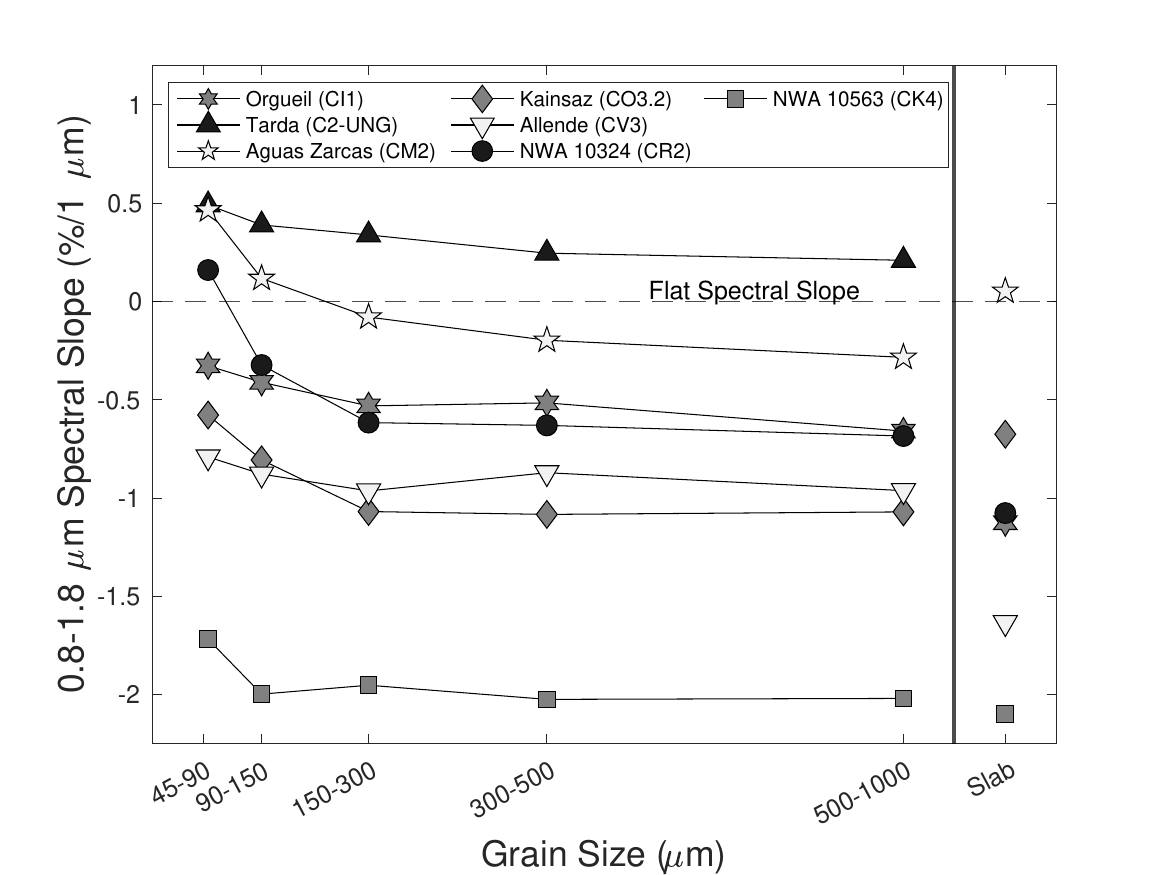}
    \caption{Plot of 0.8-1.8 $\mu$m spectral slope of our seven studied meteorite samples across our five grain sizes and slab.}
    \label{fig:slope_grainsize}
\end{figure}

\begin{figure}[H]
    \centering
    \includegraphics[width=0.8\linewidth]{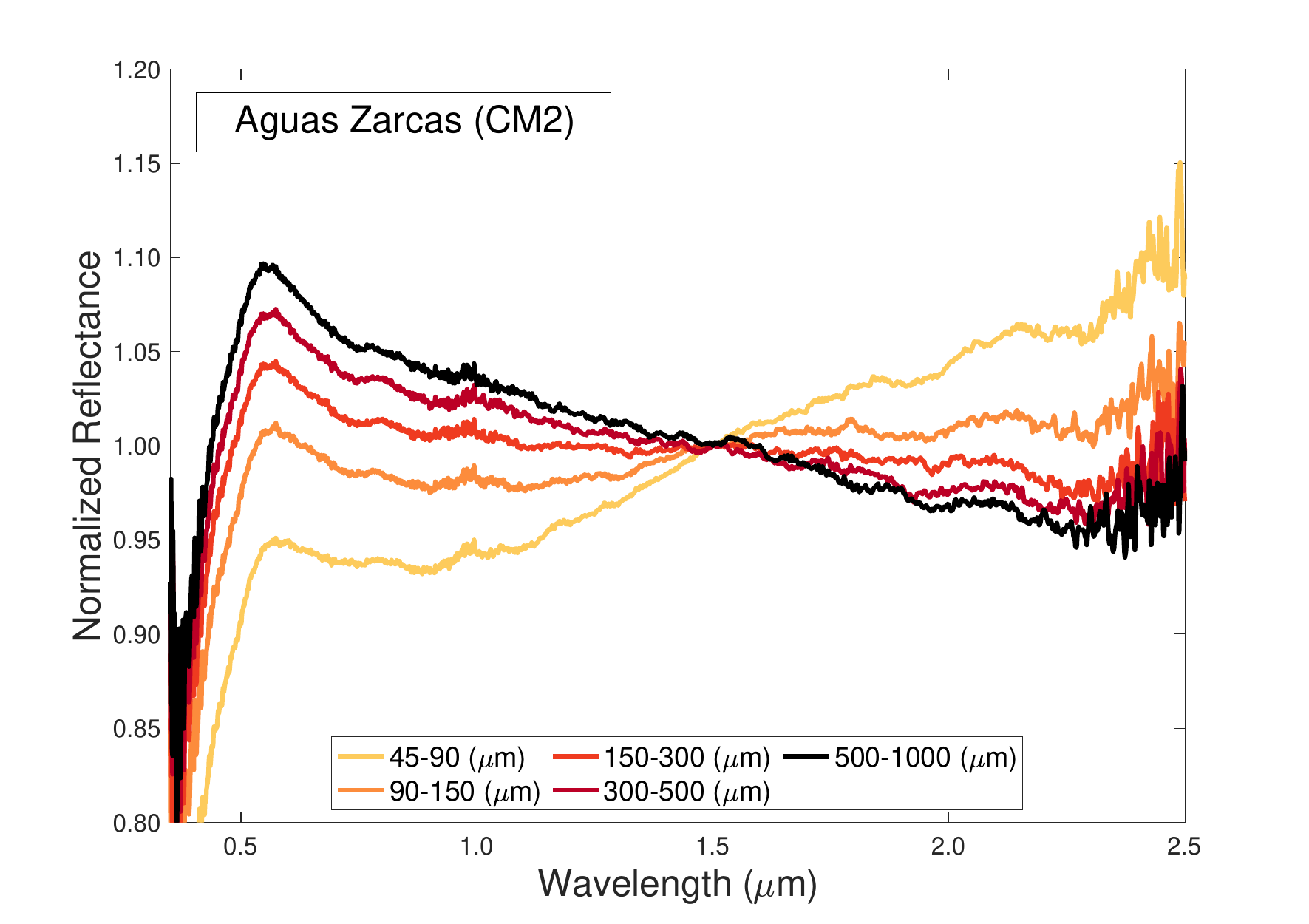}
    \caption{Plot of 0.35-2.5 $\mu$m (VNIR) spectra of CM2 Aguas Zarcas powders. Aguas Zarcas has a positive (red) spectral slope at smaller grain sizes and shifts to a negative (blue) spectral slope with increasing particle size. The data is normalized to unity at 1.5 $\mu$m.}
    \label{fig:az_grainsize}
\end{figure}

Figure \ref{fig:relative_panels} shows the normalized meteorite spectra within each column, which is useful in analyzing the spectral slope variation across grain sizes. From the data in Column 2, 0.8-1.8 $\mu$m spectral slopes are plotted in Figure \ref{fig:slope_grainsize}. All samples generally exhibit the trend of having a decreased spectral slope with increasing grain size, with some slight exceptions. Tarda has the most positive spectral slope between 0.8 and 1.8 $\mu$m and maintains this positive red slope (increasing reflectance with increasing wavelength) across the grain size bins, while Orgueil, Kainsaz, Allende, and NWA 10563 all maintain a negative blue slope (decreasing reflectance with increasing wavelength) across all grain sizes. Out of these blue-sloped meteorites, NWA 10563 has a significantly decreased spectral slope. Aguas Zarcas and NWA 10324 start with red-sloped spectra at smaller grain sizes, though shift to blue-sloped spectra with increasing grain size. For clarity, the VNIR reflectance spectra of Aguas Zarcas across its multiple grain size powders are plotted in Figure \ref{fig:az_grainsize}. While it is difficult to quantify trends using the unconstrained slab data, we can observe that most slab spectra are blue-sloped, excluding the nearly-flat slope of the Aguas Zarcas slab. \\

\begin{figure}[H]
    \centering
    \includegraphics[width=0.8\linewidth]{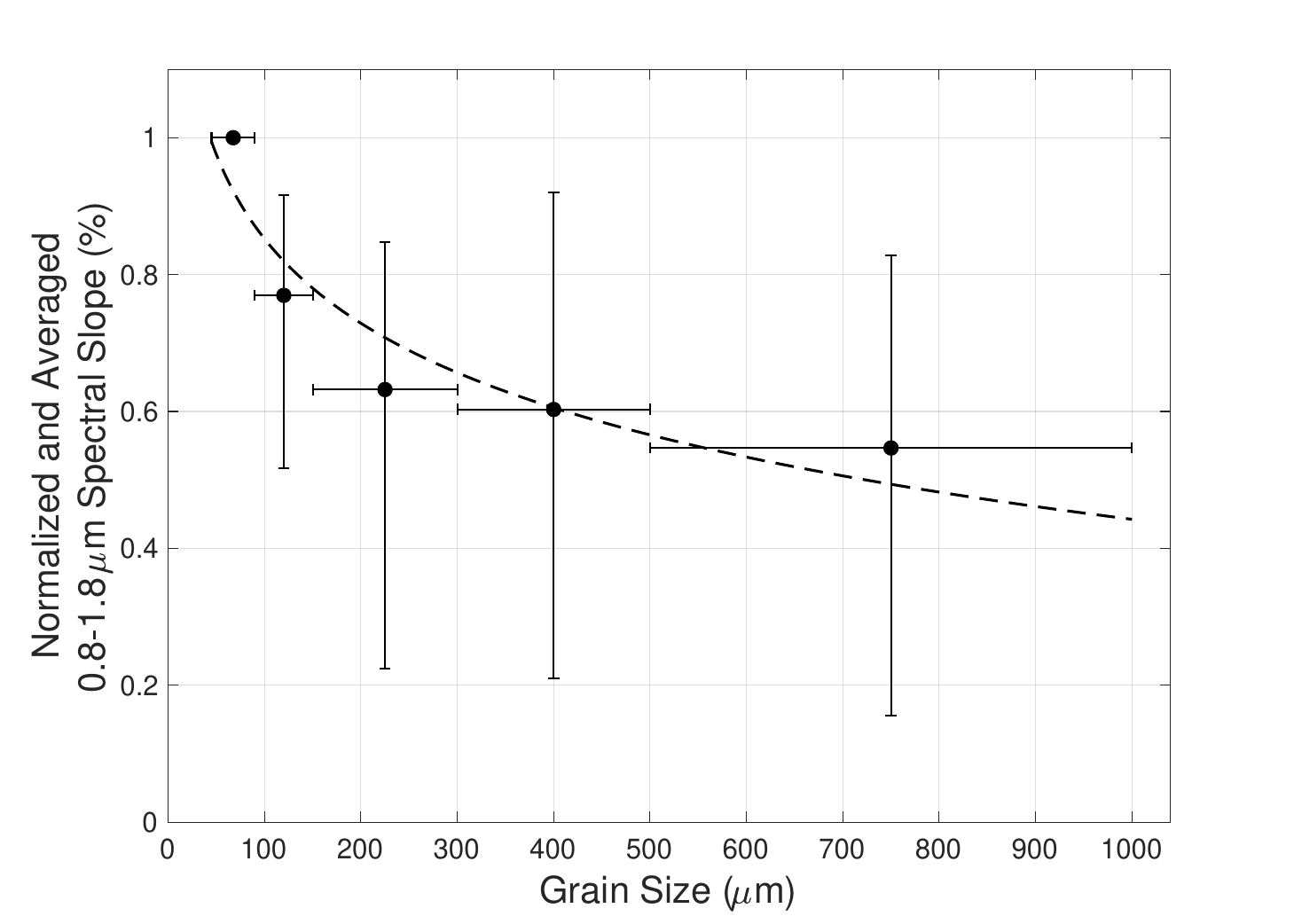}
    \caption{Plot of normalized 0.8-1.8 $\mu$m spectral slope averaged across our seven studied meteorites for each grain size. The meteorites are offset in the y-direction such that the normalized spectral slope is equal to 1.00 for the smallest grain size bin. The x-axis error bars correspond to this range in grain size between the bin's boundaries, while the y-axis error bars correspond to the minimum and maximum normalized band depth of individual meteorites in that bin. A negative natural log trend shows the best fit for the data.} 
    \label{fig:slope_trend}
\end{figure}

To better understand the relative change in slope across the different meteorites, Figure \ref{fig:slope_trend} normalizes the data from Figure \ref{fig:slope_grainsize}. From this, we can observe that NWA 10324 has the largest relative decrease in spectral slope with increasing grain size. Allende, on the other hand, shows a much more subtle change in spectral slope compared to the other meteorites. Despite the differences between meteorites, all changes in spectral slope across grain size can be reasonably modeled with a negative power law. A key takeaway is that spectral slope decreases with increasing grain size.

\subsection{Band Depths Across Grain Size}
\begin{figure}[H]
    \centering
    \includegraphics[width=0.8\linewidth]{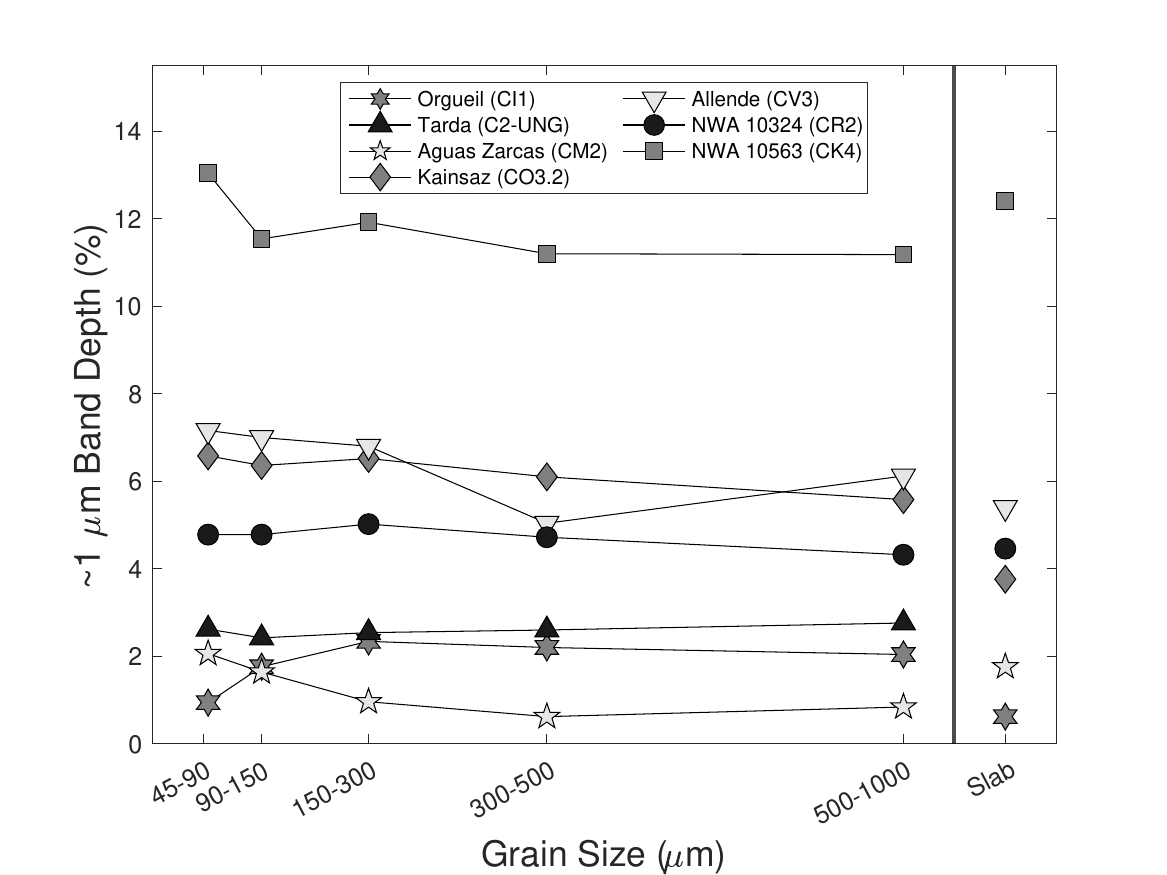}
    \caption{Plot of $\sim$1 $\mu$m band depth of our seven studied meteorite samples across our five grain size bins and slab.}
    \label{fig:depth1_grainsize}
\end{figure}

The $\sim$1.0 $\mu$m region is a widely studied feature in telescopic VNIR spectra and it is found across all of our meteorites. For Kainsaz, Allende, and NWA 10563, the 1.0 $\mu$m feature is fairly pronounced and associated mainly with olivine and pyroxene content. As petrographic type 3 and 4, these meteorites are less aqueously altered, which has allowed them to retain their largely silicate source material. Orgueil, Tarda, and Aguas Zarcas, our more aqueously alterated type 1 and 2 meteorites, lack this sharp, prominent absorption feature. Instead, their spectra are characterized by a broader magnetite absorption feature centered closer to 1.2 $\mu$m. NWA 10324, a CR2 chondrite, has a moderate 1.0 $\mu$m band depth that may be a result of only partial aqueous alteration. The apparent correlation between petrographic type and 1.0 $\mu$m band depth among carbonaceous chondrites likely stems from the suppression of silicate mineral features and their chemical alteration into phyllosilicates that lack this distinct 1.0 $\mu$m feature. To understand the change in this band depth across grain size, we measured the strongest band depth from 0.9-1.3 $\mu$m for each meteorite. Acknowledging that this feature may be a result of slightly different mineralogies across each meteorite, we included the band center values in Tables \ref{table:band_parameters_1} and \ref{table:band_parameters_2}. The band depths of this $\sim$1.0 $\mu$m feature across grain size for each of our meteorites are shown in Figure \ref{fig:depth1_grainsize}. Trends in 1.0 $\mu$m band depth versus grain size are difficult to discern. NWA 10563 and Aguas Zarcas appear to subtly decrease in 1.0 $\mu$m band depth with larger grain sizes, with the 1.0 $\mu$m band depth of Aguas Zarcas nearly disappearing altogether with larger grain sizes. We suspect that this is due to less scattering of light caused by the decreased optical path length of our larger-grained samples. As a result, there is generally a reduction in absorption band depths with increasing grain size. Orgueil uniquely shows the opposite trend of increasing 1.0 $\mu$m band depth with increasing grain size until $\sim$300 $\mu$m, after which it remains flat. All other meteorites show little to no discernible trend in 1.0 $\mu$m band depth across grain size.

\subsection{Band Centers Across Grain Size}
\begin{figure}[H]
    \centering
    \includegraphics[width=0.8\linewidth]{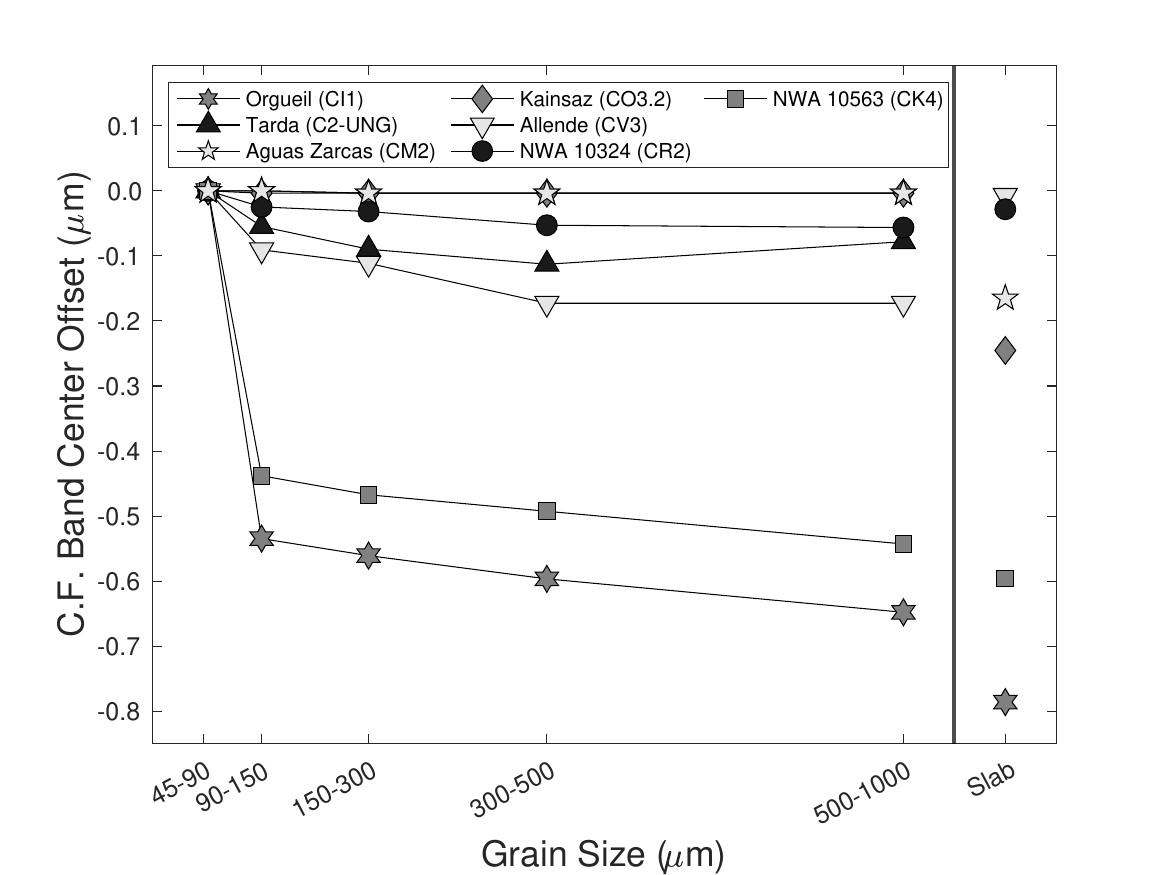}
    \caption{Plot of the offset in Christiansen feature band center relative to the 45-90 $\mu$m powder for each of our seven studied meteorite samples across our five grain size bins and slab.}
    \label{fig:CF_grainsize}
\end{figure}

\begin{figure}[H]
    \centering
    \includegraphics[width=0.8\linewidth]{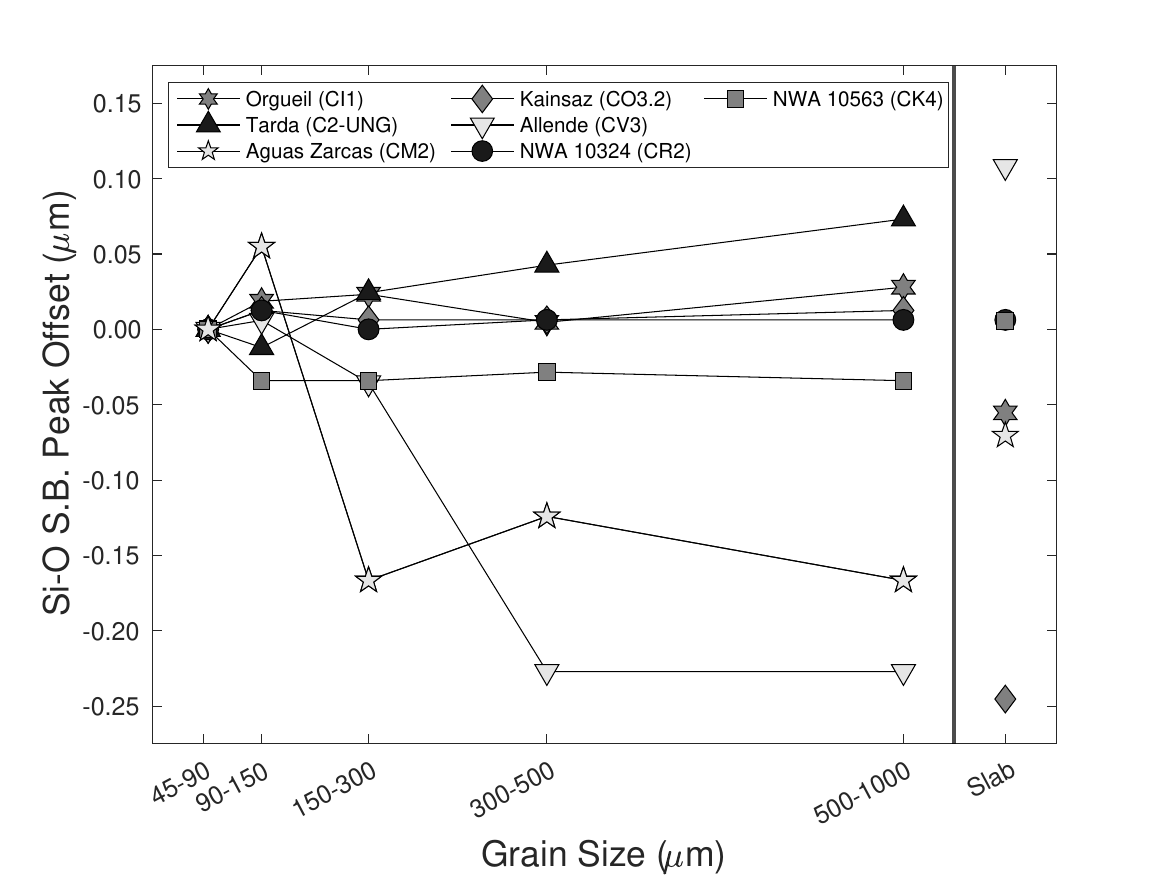}
    \caption{Plot of the offset in 9.5-11.5 $\mu$m Si-O stretching band peak center relative to the 45-90 $\mu$m powder for each of our seven studied meteorite samples across our five grain size bins and slab.}
    \label{fig:Si-O_grainsize}
\end{figure}

For our band center analysis, we chose to study the C.F. minima and Si-O S.B. maxima centers as a function of grain size, as these metrics can be used to estimate the degree of aqueous alteration in carbonaceous chondrite meteorites \citep{beck2014transmission,mcadam2015aqueous,bates2020linking,hanna2020distinguishing}. The C.F. minimum and Si-O S.B. peak locations for our 45-90 $\mu$m meteorite samples are listed in Tables \ref{table:band_parameters_1} and \ref{table:band_parameters_2}. From here, the wavelength shift of these features from this 45-90 $\mu$m value for each of our samples is plotted in Figures \ref{fig:CF_grainsize} and \ref{fig:Si-O_grainsize}. For the C.F. in Figure \ref{fig:CF_grainsize}, we observe a general negative offset, or shorter wavelength shift in C.F., with increasing grain size. This is most obvious in Orgueil, NWA 10563, and Allende, with wavelength offsets of up to -0.65 $\mu$m observed in the 500-1000 $\mu$m Orgueil powder (see Figures \ref{fig:absolute_panels} and \ref{fig:relative_panels}). Aguas Zarcas and Kainsaz do not show any discernible shift in C.F. as a function of grain size, not including their slab data that shows a shorter wavelength shift. None of our samples exhibit a positive, or longer wavelength C.F. shift with increasing grain size. In Figure \ref{fig:Si-O_grainsize}, we find less of a general trend in the Si-O S.B. peak center across grain size. Allende has a shorter wave Si-O S.B. peak center with increasing grain size, while Tarda displays the opposite trend with increasing grain size. It is worth noting that with increasing grain size, certain meteorites displayed different Si-O S.B. peaks than what was originally observed in their 45-90 $\mu$m sample. One good example of this is Tarda, which has a peak position at 11.21 $\mu$m in its 45-90 $\mu$m sample. With increasing grain size, this feature remains present but is only a relative peak compared to a growing 9.82 $\mu$m feature. To keep consistency in our plot, we only tracked the peak position that was largest in the 45-90 $\mu$m sample, whether it stayed as a maximum or relative peak center with increasing grain size. While there is less of a trend than what is observed with the C.F., there is moderate variability in the Si-O S.B. peak across grain size, with offset from the 45-90 $\mu$m values of up to 0.25 $\mu$m. As mentioned in the Methodology section, it is possible that the presence of hyperfine grains may also subtly influence these mid-infrared features.
%} \\

While the C.F. and Si-O S.B. positions can individually provide a good gauge on aqueous alteration, \citet{hanna2020distinguishing} found that the separation between the C.F. and Si-O S.B. positions provides the best correlation to aqueous alteration in CM2 meteorites. To compare our results, we calculated the predicted petrologic subtype for our CM2 chondrite Aguas Zarcas given the equation provided by \citet{hanna2020distinguishing} and using the \citet{rubin2007progressive} alteration scale. The predicted subtype across all grain sizes generally falls in the 2.0 range, which is expected for a type 2 meteorite. Across different grain sizes, the predicted petrologic subtype varies by up to 0.25, stemming from Aguas Zarcas' shift in Si-O S.B. center, but no obvious trend is observed.

\begin{table}[H]
    \centering
    \caption{\textbf{Predicted Petrologic Subtype of Aguas Zarcas.}}
    \begin{tabular}{|c|c|}
        \hline
        \textbf{Grain Size} & \textbf{Petrologic Subtype} \\
        \hline
        45-90 $\mu$m & 2.57 \\
        \hline
        90-150 $\mu$m & 2.60 \\
        \hline
        150-300 $\mu$m & 2.45 \\
        \hline
        300-500 $\mu$m & 2.48 \\
        \hline
        500-1000 $\mu$m & 2.45 \\
        \hline
        Slab & 2.70 \\
        \hline
    \end{tabular}
    % \captionsetup{width=0.75\textwidth, justification=justified, singlelinecheck=false}
    \tablecomments{Predicted petrologic subtype was found using the C.F. and Si-O S.B. separation equation from \citet{hanna2020distinguishing} with the \citet{rubin2007progressive} alteration scale.}
\end{table}

\subsection{Summary of Results}
We observed general trends in spectral parameters across grain size for our seven meteorites of study. Despite being different carbonaceous chondrite types, our samples showed nearly identical band features in the UV-MIR (0.2-14 $\mu$m) that varied in depth and center. Kainsaz, NWA 10324, and NWA 10563 displayed more prominent weathering absorption features near 0.50 $\mu$m, while Aguas Zarcas was the only meteorite displaying a 0.70 $\mu$m band feature relating to hydrated phyllosilicates. All meteorites showed signs of a $\sim$1.0 absorption feature, with additional bands present in the infrared for only select meteorites. From 0.2-8.0 $\mu$m, our meteorites show the general trend of decreased absolute reflectance with increasing grain size. Past $\sim$8.0 $\mu$m, this relationship with grain size and reflectance is more ambiguous. In the near-infrared (0.8-1.8 $\mu$m), we observe a decreased spectral slope with increasing grain size. Our band depth vs. grain size analysis yielded an interesting result: trends in the 1.0 $\mu$m area are difficult to distinguish but show absolute differences across meteorites based on petrographic type. Band centers between each of the meteorites vary slightly due to compositional differences, though variation across grain size is observed in the shifting of the C.F. towards shorter wavelengths. Moderate variation is observed in the shifting of the Si-O S.B. peak across grain size, but no clear trends are observed. \\

While each meteorite varied in reflectance across grain size and wavelength, it is possible to broadly group them together. On the most basic level, Orgueil, Tarda, and Aguas Zarcas behaved the most similar with low albedos and weak 1.0 $\mu$m band depths. Their similarities may lie in their composition and lack of significant terrestrial weathering. Kainsaz, NWA 10324, and NWA 10563 can also be grouped as having similar trends with increasing grain size. Our last meteorite, Allende, behaved the most differently from our other samples and did not usually follow the trends with grain size.

\section{Analysis}
\label{analysis}

\subsection{Principal Component Analysis}
Principal component analysis (PCA) is a dimensionality reduction tool that can help better understand complex data sets. The original asteroid taxonomy system established by \citet{tholen1984asteroid} and all subsequent major asteroid taxonomy systems were based on PCA \citep{Bus1999compositional,bus2002phase,bus2002phaseb,demeo2009extension}. PCA can be used to provide some insight into an asteroid’s taxonomy but does not provide rigorous surface mineralogy information. PC values were measured using the online tool from the MIT Small Main-Belt Asteroid Spectroscopic Survey (SMASS).

\begin{figure}[H]
    \centering
    \includegraphics[width=0.8\linewidth]{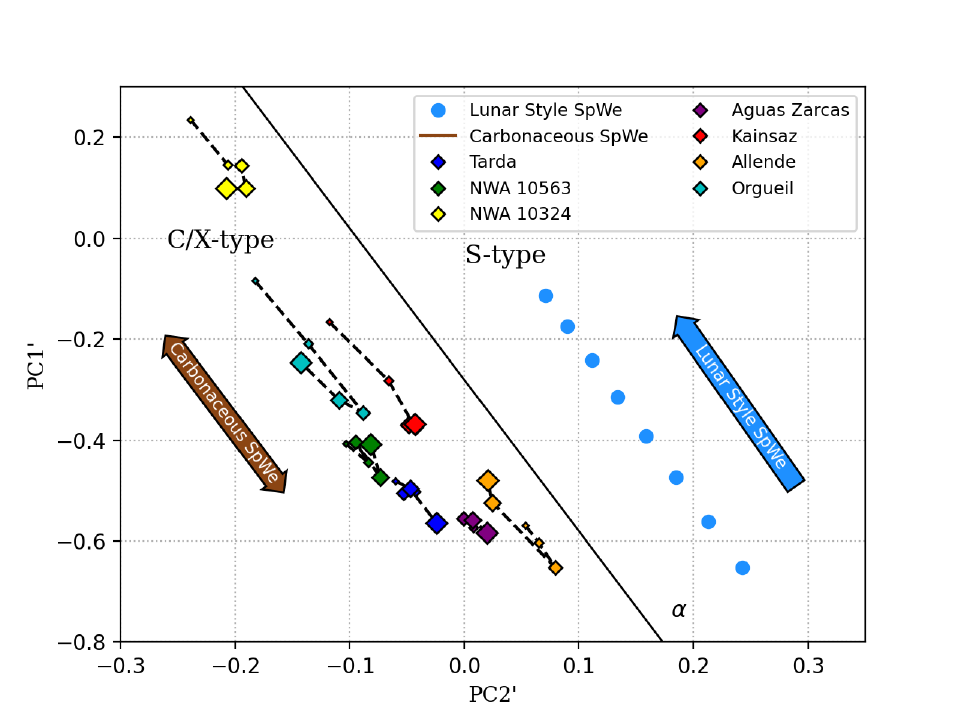}
    \caption{Principal component values for all six meteorites studied at different grain sizes are shown with the “$\alpha$-line” boundary from \citet{demeo2009extension} that separates the S- and C/X-complexes. Grain size variations can be seen by using the dotted line interconnecting PC values of measurements of the same meteorite at different grain sizes. Marker size increases with grain size which includes values of 45-90 $\mu$m, 90-150 $\mu$m, 150-300 $\mu$m, 300-500 $\mu$m, and 500-1,000 $\mu$m. The Lunar-style space weathering trend from \citet{binzel2019compositional} is shown with blue dots and a corresponding arrow showing the direction of increasing space weathering. The carbonaceous space weathering trend from \citet{lantz2018space} is shown with the brown arrow indicating that the measured weathering trend would move in either direction depending on the meteorite composition. It also shows that differentiating between space weathering and grain size effects would be challenging in PC space.}
    \label{fig:PCA}
\end{figure}

PC values were measured for all six meteorites at grain sizes of 45-90 $\mu$m, 90-150 $\mu$m, 150-300 $\mu$m, 300-500 $\mu$m, and 500-1,000 $\mu$m, and the results are plotted in Figure \ref{fig:PCA}. Many of the meteorites show relatively large (0.1 – 0.2) variations in their PC values across the different grain sizes with Aguas Zarcas as the exception since its PC values are clustered around a single point. The most notable trend across all meteorites is that the changes in PC values owing to grain size variations are predominantly parallel to the $\alpha$-line that separates typically feature-rich S-type spectra from typically featureless C/X-type spectra.

\subsection{Bus-DeMeo Taxonomic Classification}
Using the same online Bus-DeMeo asteroid taxonomic tool, classifications were found for each meteorite's available spectra and plotted in Table \ref{table:asteroidtaxonomy}. CI1 Orgueil and CV3 Allende are classified as B-types across all grain sizes, while CR2 NWA 10324 and CK4 NWA 10563 are classified as an L-type and K-type across all grain sizes, respectively. C2-UNG Tarda is largely classified as a C-type asteroid from 45-500 $\mu$m, though shifts to a Cb-type that indicates a positive slope forming near 1.1 $\mu$m instead of 1.3 $\mu$m associated with C-types \citep{demeo2009extension}. CM2 Aguas Zarcas also shows a shift in classification with increasing grain size, transforming from a Ch to B-type classification. This change in classification is likely a result of the decreasing spectral slope seen in Figure \ref{fig:slope_grainsize}. Lastly, CO3.2 Kainsaz has a different slab classification (B-type) than its powdered samples (K-type), though it is difficult to interpret the effects of grain size as slab samples are unconstrained in grain size.

\begin{table}[H]
    \centering
    \caption{Bus-DeMeo Asteroid Taxonomic Classification}
    \label{tb5}
    \addtolength{\tabcolsep}{-13pt}
    \begin{tabular}{lccccccc}

        \textbf{Sample Name} &
        \scell[c]{c}{\textbf{Type}} &
        \scell[c]{c}{\textbf{45-90 $\mu$m}} & 
        \scell[c]{c}{\textbf{90-150 $\mu$m}} & 
        \scell[c]{c}{\textbf{150-300 $\mu$m}} & 
        \scell[c]{c}{\textbf{300-500 $\mu$m}} & 
        \scell[c]{c}{\textbf{500-1000 $\mu$m}} &
        \scell[c]{c}{\textbf{Slab}} \\

        \noalign{\smallskip}\hline\hline
        Orgueil & CI1 & B & B & B & B & B & B\\
        
        \hline\noalign{\smallskip}
        
        Tarda & C2-UNG & C & C & C & 
        C & Cb & \\

        \hline\noalign{\smallskip}

        Aguas Zarcas & CM2 & Ch & Ch & B & B & B
        & B \\

        \hline\noalign{\smallskip}

        Kainsaz & CO3.2 & K & K & K & K & K & B \\

        \hline\noalign{\smallskip}

        Allende & CV3 & B & B & B & B & B & B\\

        \hline\noalign{\smallskip}

        NWA 10324 & CR2 & L & L & L & L & L & L \\

        \hline\noalign{\smallskip}

        NWA 10563 & CK4 & K & K & K & K & K & K \\

        \noalign{\smallskip}\hline
    \end{tabular}

    % \captionsetup{width=0.7\textwidth, justification=justified, 
    %               singlelinecheck=false}
    \tablecomments{Asteroid taxonomy classification performed using a Bus-DeMeo Taxonomy Classification Web tool. Principal component analysis detailed by \citet{demeo2009extension}. }
    \label{table:asteroidtaxonomy}
\end{table}

\subsection{Hapke Model and Its Parameterization for Meteorites}

Bidirectional reflectance is often employed in several real-world applications such as computer graphics for photo-realistic rendering of synthetic scenes and in computer vision for object recognition \citep{marschner2000image,dana2004device}, but its greatest use is in remote sensing for surface reflectance characterization \citep{roujean1992bidirectional}. To model the reflectance of a material in a given wavelength as a function of the Single Scattering Albedo (SSA) and scattering parameters, and of the geometry of the experiment setup for data acquisition \citep{mustard1989photometric}, the Hapke model \citep{hapke2012theory} is an effective tool that can be employed. The Hapke model considered for this work \citep{hapke1981bidirectional} is governed by four main components. This first is the Lommel-Seeliger reflection coefficient, which includes the single-scattering albedo $\omega \in [0,1]$ (SSA), and the cosines of the emission $(e)$ and incidence $(i)$ angles, $\mu=\cos e$, and $\mu_0 = \cos i$, respectively. The second component is the particle phase function $P(g)$, which can be expressed by the double Henyey-Greenstein (HG) function \citep{hapke2012theory}, depending on the phase angle $g$ (given by the angle between the direction of incident light and the direction of the reflected light) and the scattering parameters $b$ and $c$, describing the scattering behavior of the surface. The parameter $b\in [0,1]$ represents the angular width of the scattering lobe. When the value of $b$ is large, the scattering lobe is tall and narrow, and vice versa for a smaller value of $b$. The parameter $c$, representing the amplitude of the back-scattered lobe relative to the forward lobe, can assume positive and negative values in the range $[-1,1]$. When $c$ is negative, a forward scattering is predominant, while when $c$ assumes positive values, we can expect to have mainly backward scattering \citep{pilorget2015photometry}. The third component is represented by the opposition surge function $B(g)$, which for the current experiments we can neglect since the phase angle $g\geq 15^o$ \citep{mustard1989photometric}. The fourth and last Hapke model component is Chandrasekhar’s isotropic scattering H-function $H(\omega,\chi)$, a function of the SSA $\omega$ and of the cosine of the incidence/emission angle \citep{chandrasekhar2013radiative}.\\

Hence, the Hapke model we adopt has the following form

\begin{equation}
    R(\lambda) = \frac{\omega}{4(\mu - \mu_0)} \left[ P(g) + H(\omega,\mu)H(\omega,\mu_0) - 1 \right]
\end{equation}

The function $R(\lambda)$ is the bidirectional radiance coefficient, and it represents the brightness of a surface relative to the brightness of a Lambert surface identically illuminated \citep{hapke1981bidirectional}. As H-function, we use an approximation with relative errors $< 1\%$ proposed in Ref. \citep{hapke2012theory}, that is

\begin{equation}
    H(\omega,\mu) = \left\{  1 - (1 - \sqrt{1 -\omega}) \mu \left[  r_0 + \left( 1 - \frac{1}{2} r_0 \mu \right) \ln{ \left(  \frac{1+\mu}{\mu} \right)}      \right]     \right\}^{-1}
\end{equation}

\noindent with the diffusive reflectance $r_0$ expressed as

\begin{equation}
    r_0 = \frac{1 - \sqrt{1 - \omega}}{1 + \sqrt{1 - \omega}}
\end{equation}

Depending on the assumption of the scattering we are facing, we must treat the phase function $P(g)$ accordingly. If the scattering is assumed to be isotropic, $P(g)$ is simply set equal to 1. While if an anisotropic scattering is considered, the single-scattering phase function $P(g)$ can be modeled by the double HG function \citep{hapke2012bidirectional} as follows

\begin{equation}
    P(g) = \frac{1+c}{2} P_{HGB}(g,b) + \frac{1-c}{2} P_{HGF}(g,b)
\end{equation}

\noindent where $P_{HGB}$ and $P_{HGF}$ represent the backward-scattered lobe centered in $g=0^o$ and the forward-scattered lobe centered in $g=180^o$, respectively, and can be expressed as

\begin{equation}
    P_{HGB}(g,b) = \frac{1-b^2}{ (1 - 2b \cos g + b^2)^{3/2} } \qquad \text{and} \qquad P_{HGF}(g,b) = \frac{1-b^2}{ (1 + 2b \cos g + b^2)^{3/2} }
\end{equation}

Thus, we can write the Hapke model formulation for isotropic scattering containing the SSA $\omega$ as 

\begin{equation}
    R_{iso}(\lambda) = \frac{\omega}{4(\mu - \mu_0)} \cdot    \frac{ \left\{  1 - (1 - \sqrt{1 -\omega}) \mu \left[  \frac{1 - \sqrt{1 - \omega}}{1 + \sqrt{1 - \omega}} + \left( 1 - \frac{1}{2} \frac{1 - \sqrt{1 - \omega}}{1 + \sqrt{1 - \omega}} \mu \right) \ln{ \left(  \frac{1+\mu}{\mu} \right)}      \right]     \right\}^{-1}   }{   1 - (1 - \sqrt{1 -\omega}) \mu \left[  \frac{1 - \sqrt{1 - \omega}}{1 + \sqrt{1 - \omega}} + \left( 1 - \frac{1}{2} \frac{1 - \sqrt{1 - \omega}}{1 + \sqrt{1 - \omega}} \mu \right) \ln{ \left(  \frac{1+\mu}{\mu} \right)}      \right]  }  
\end{equation}

\noindent and the Hapke model for anisotropic scattering containing $\omega$, $b$, and $c$, as

\begin{multline}
    R_{aniso}(\lambda) = \frac{\omega}{4(\mu - \mu_0)} \cdot \Bigg\{ \frac{1+c}{2} \frac{1-b^2}{ (1 - 2b \cos g + b^2)^{3/2} } +  \frac{1-c}{2}  \frac{1-b^2}{ (1 + 2b \cos g + b^2)^{3/2} }   \\
    + \frac{ \left\{  1 - (1 - \sqrt{1 -\omega}) \mu \left[  \frac{1 - \sqrt{1 - \omega}}{1 + \sqrt{1 - \omega}} + \left( 1 - \frac{1}{2} \frac{1 - \sqrt{1 - \omega}}{1 + \sqrt{1 - \omega}} \mu \right) \ln{ \left(  \frac{1+\mu}{\mu} \right)}      \right]     \right\}^{-1}   }{   1 - (1 - \sqrt{1 -\omega}) \mu \left[  \frac{1 - \sqrt{1 - \omega}}{1 + \sqrt{1 - \omega}} + \left( 1 - \frac{1}{2} \frac{1 - \sqrt{1 - \omega}}{1 + \sqrt{1 - \omega}} \mu \right) \ln{ \left(  \frac{1+\mu}{\mu} \right)}      \right]  }- 1   \Bigg\}
\end{multline}

The optimization problem we aim to solve consists of determining the optimal set of variables ($\omega$ for the isotropic case and $(\omega,b,c)$ for the anisotropic case) that minimizes the error between observed reflectance and modeled reflectance for the geometry setup used in our measurements. The functions we want to minimize are the chi-square values

\begin{equation}
    \chi_{iso}^2  \left[ \frac{R_{obs}(i,e,g) - R_{modeled}(i,e,g,\omega)}{\sigma}   \right]^2 \qquad \text{and} \qquad  \chi_{aniso}^2  \left[ \frac{R_{obs}(i,e,g) - R_{modeled}(i,e,g,\omega,b,c)}{\sigma}   \right]^2
\end{equation}

\noindent where $R_{obs}$ represents the reflectance measured in the laboratory, $R_{modeled}$ represents the reflectance obtained by the optimization model, and $\sigma$ is the standard deviation of $R_{obs}$. \\

The optimization is carried out by using the function \textit{fmincon} implemented in MATLAB to exploit the interior-point algorithm and compute the set of optimization variables $(\omega,b,c)$. The variable values are constrained in the ranges of values mentioned above. The first initial value guess in the algorithm has been set equal to $[1,0,1]$, and once the optimal set $(\omega,b,c)$ for the first wavelength is computed, it is subsequently used as an initial value guess for the next wavelength.

\subsubsection{Grain Size Regression Model}

For each meteorite sample, we retrieved a set of photometric Hapke model parameters as the first step. To study the relationship between the photometric parameters (scattering parameters) with grain size, we set up different linear regression models to focus on the evolution of the parameter $\omega$. For particle size, we used five values taken as the average of the minimum and maximum grain sizes for each grain size bin. To evaluate the retrieved grain size, three different regression models are used, such as polynomial curve fitting \citep{eubank1990curve}, a modified Akima interpolation \citep{akima1991method}, and a Gaussian process regression \citep{quinonero2005unifying}. The regressions have been applied to both the isotropic and the anisotropic cases.

\subsubsection{Derivation of the Hapke Model parameters and Empirical Relationship Between the SSA and Particle Size}

\begin{figure}[H]
    \centering
    \includegraphics[width=0.8\linewidth]{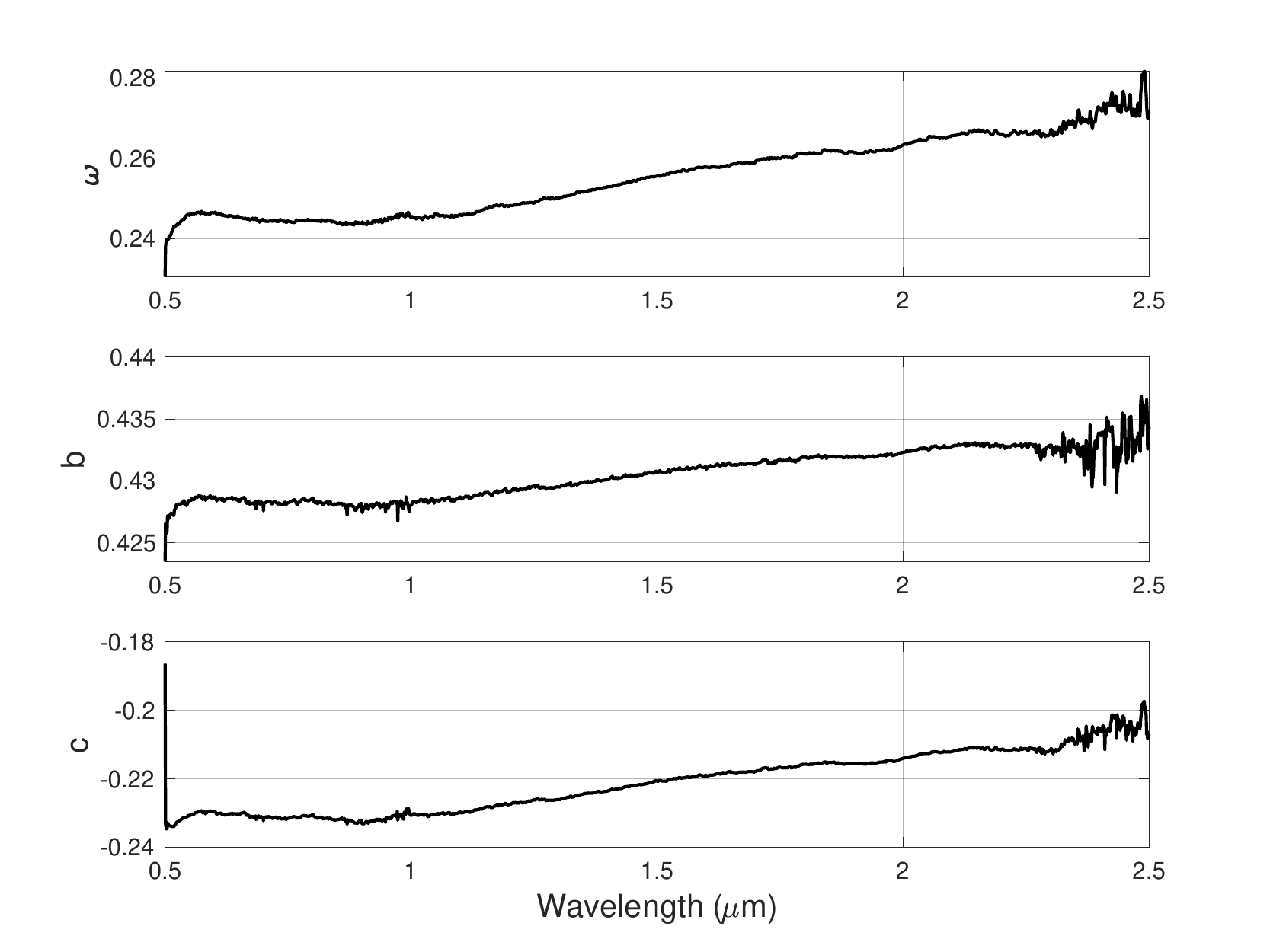}
    \caption{Optimal set of variables for anisotropic scattering $\omega$, $b$, and $c$ for Aguas Zarcas $45-90 \mu$m from our modeling effort.}
    \label{fig:wbc4590}
\end{figure}

For the sake of simplicity, here we report only the Hapke model output obtained with one of our seven meteorites, Aguas Zarcas, sampled in the NIR region (0.5-2.5 $\mu$m). In Figure \ref{fig:wbc4590} the optimal set of variables $\omega$, $b$, and $c$ are shown, as functions of the wavelength $\lambda$, for the Aguas Zarcas sample $45-90 \mu$m. The single scattering albedo $\omega$ has values ranging from 0.24 to 0.28. The average value of the parameter $b$ is 4.3, which suggests a slightly short and wide scattering lobe, and the parameter $c$ assumes negative values, which shows a predominantly forward scattering behavior.  \\
To evaluate the reliability of our modeling, the left plot in Figure \ref{fig:refobs_mod_4590} shows the Reflectance measured in the laboratory versus the reflectance computed via the Hapke model. One can qualitatively see the overlapping of the two functions, which suggests the high accuracy of the Hapke model we chose in representing the meteorite's reflectance. The right plot gives us a quantitative evaluation of the precision of our Hapke model, with average values of chi-square of about $10^{-13}$.

\begin{figure}[H]
    \centering
    \includegraphics[width=0.9\linewidth]{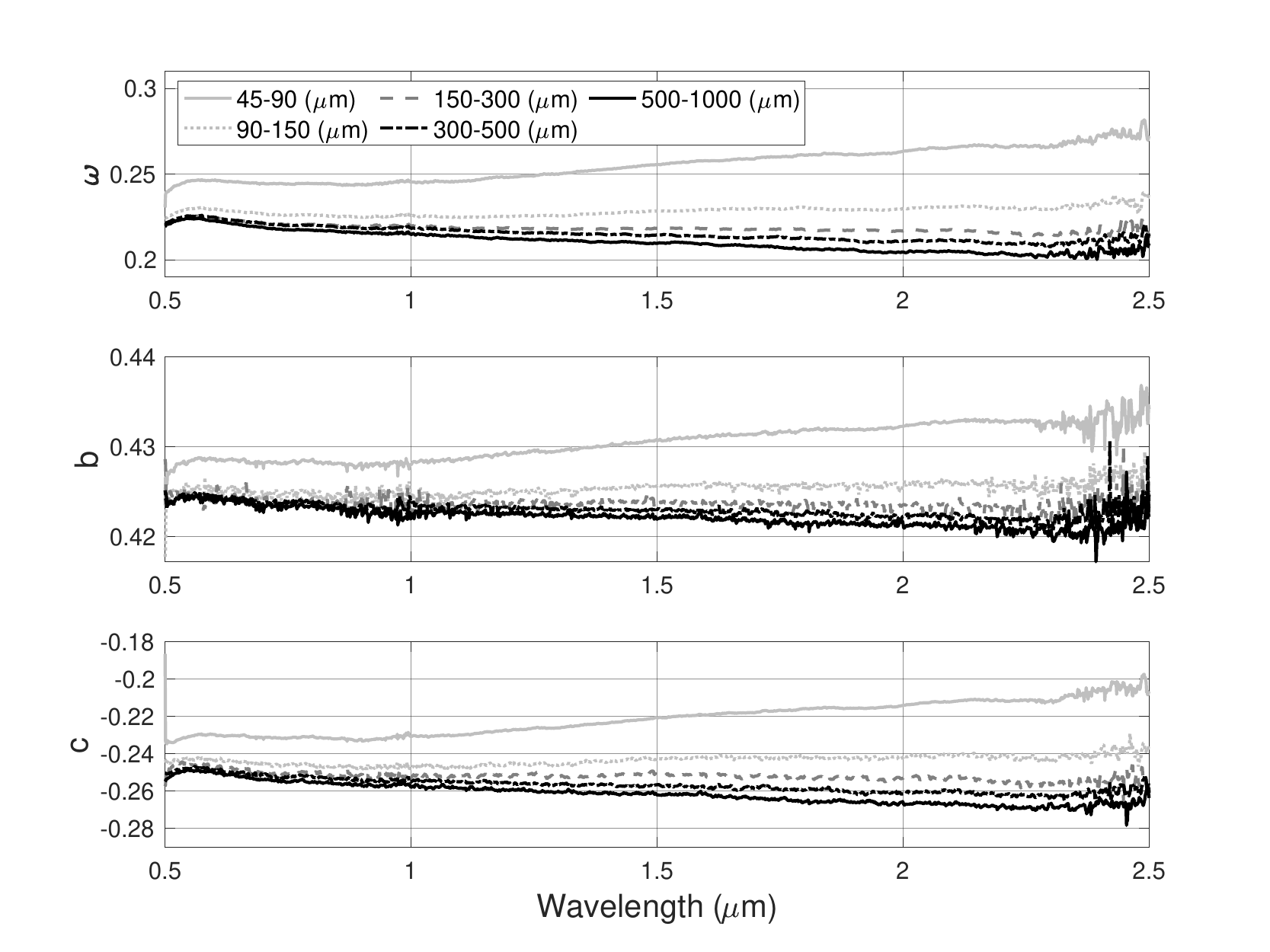}
    \caption{Optimal set of variables for anisotropic scattering $\omega$, $b$, and $c$ for all grain sizes of Aguas Zarcas.}
    \label{fig:w_b_c_total}
\end{figure}

\begin{figure}
  \centering
  \begin{tabular}[b]{@{}p{0.7\textwidth}@{}}
    \centering\includegraphics[width=\linewidth]{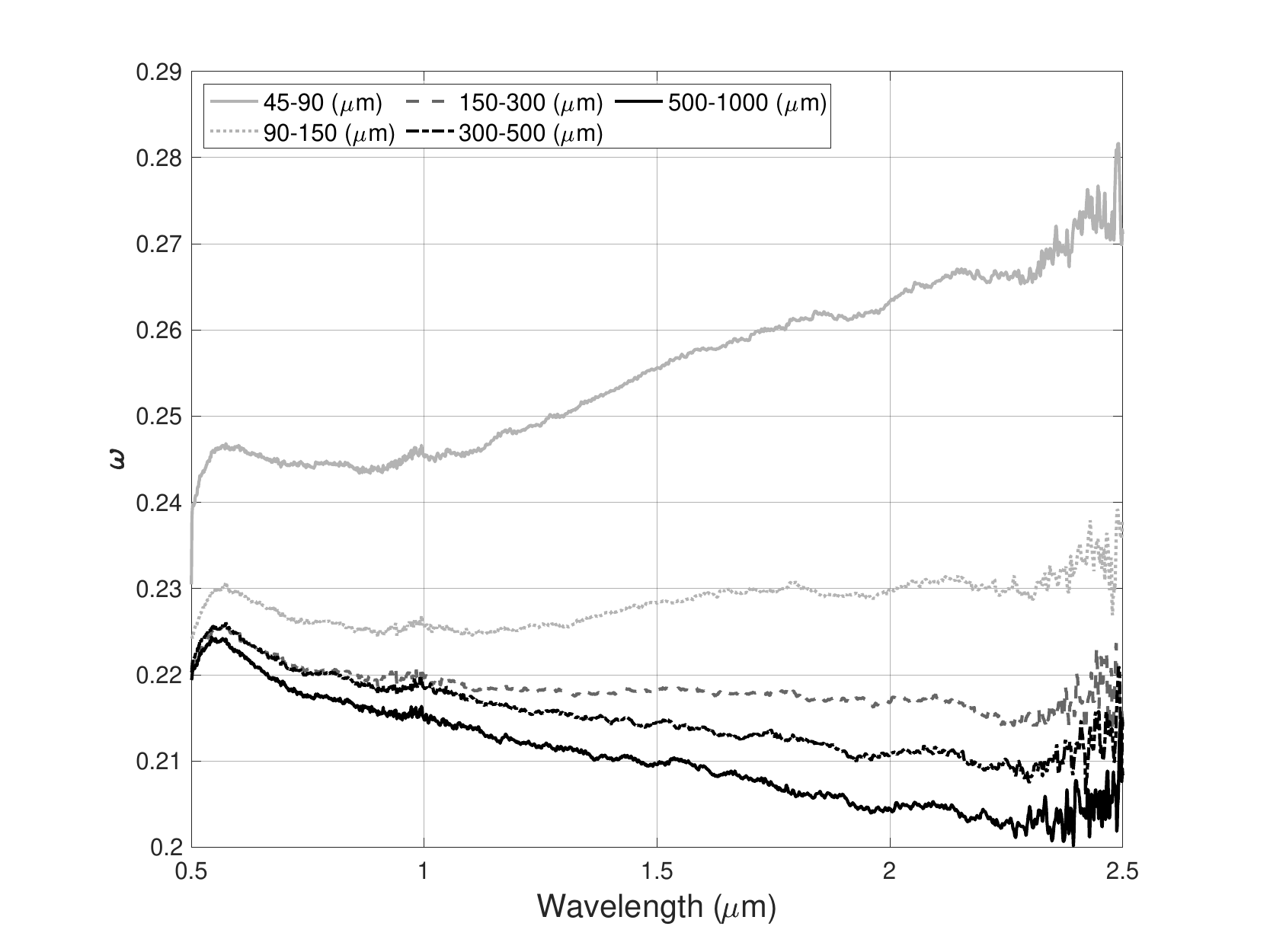} \\
    \centering\small (a) Optimal variable $\omega$ for anisotropic scattering
  \end{tabular}%
  \quad
  \begin{tabular}[b]{@{}p{0.7\textwidth}@{}}
    \centering\includegraphics[width=\linewidth]{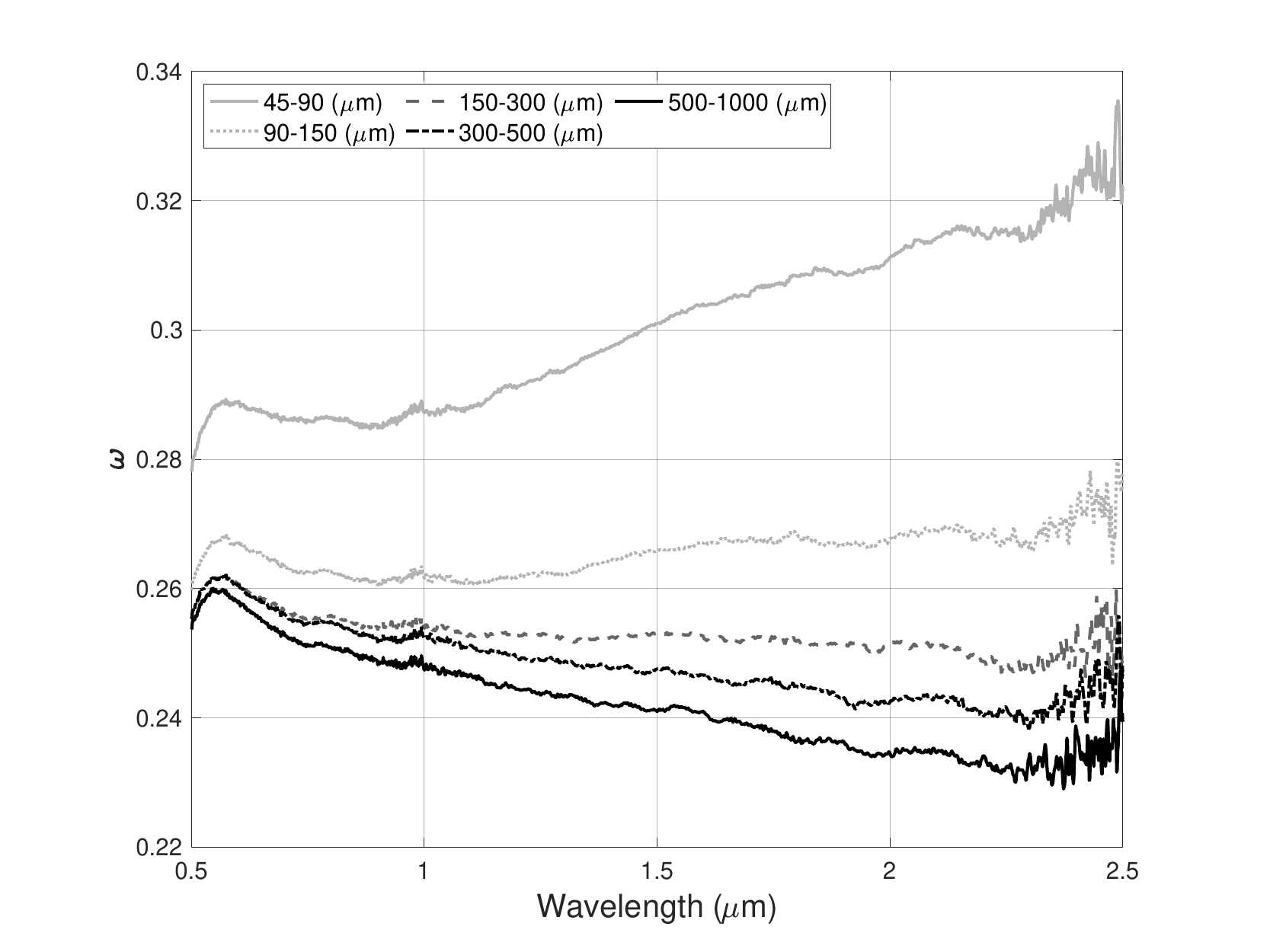} \\
    \centering\small (b) Optimal variable $\omega$ for isotropic scattering
  \end{tabular}
  \caption{Optimal variable $\omega$ for anisotropic scattering (a) and for isotropic scattering (b), for all grain sizes of Aguas Zarcas.}
  \label{fig:w}
\end{figure}

Finally, the set of optimal variables $\omega$, $b$, and $c$ for Aguas Zarcas is plotted along the wavelength in Figure \ref{fig:w_b_c_total}, for each grain size in the anisotropic case. The parameter $b$ ranges from 0.42 to 0.435 for all the samples, in an almost constant trend, exhibiting a weak dependence on the wavelength. The parameter $c$ ranges from -0.26 to -0.2, which as stated above, represents a predominantly forward scattering behavior for all grain sizes. The estimated $\omega$ varies between 0.2 to 0.28 over the NIR wavelength range. Its evolution agrees with the reflectance spectrum, and as expected, the values of $\omega$ increase with the decreasing of the grain size.\\
In Figure \ref{fig:w} the evolution of the fitted $\omega$ values for both anisotropic (left) and isotropic (right) scattering is plotted. In both cases, we have a decreased value of $\omega$ with increasing grain size, but we can note that the $\omega$ values for the simplified isotropic case are larger than those for the anisotropic case. This is consistent with previous observations of the $c$ negative values, that is, a mainly forward scattering corresponds to a smaller amount of reflected light from the surface back to the camera (smaller $\omega$).

\begin{figure}[H]
    \centering
    \includegraphics[width=1.1\linewidth]{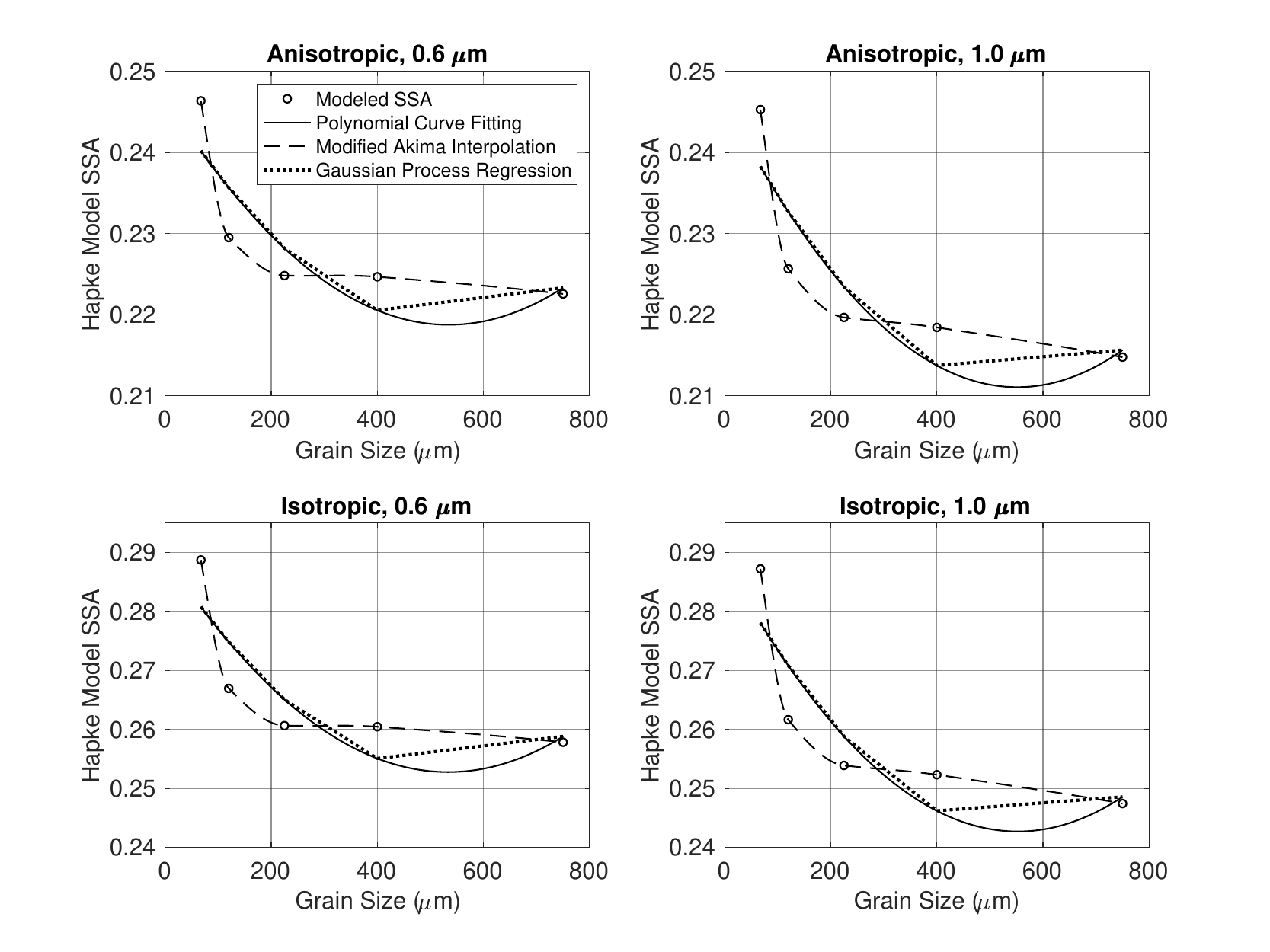}
    \caption{Regression models for the simulated single scattering albedo for anisotropic scattering (top) and isotropic scattering (bottom).}
    \label{fig:retrieval}
\end{figure}

The simulated single scattering albedos $\omega$ from the Hapke model for the anisotropic scattering (top) and isotropic scattering (bottom) are plotted in Figure \ref{fig:retrieval} against the grain sizes at 0.6 and 1.0 $\mu$m. We report the three regression models for both scattering cases and for both the chosen wavelengths. All these models suggest an asymptotic relationship between the SSA that decreases with increasing grain size. The equations for constraining the Aguas Zarcas grain size using SSA for the wavelengths 0.6 and 1.0 microns are, respectively:
\begin{equation}
    \text{grain size}\,[\mu m] = \frac{1.0518\times10^{-4} \pm \sqrt{3.9360\times10^{-7} \omega - 8.6116\times10^{-8}}}{1.9680\times10^{-7}} 
\end{equation}
and
\begin{equation}
    \text{grain size}\,[\mu m] = \frac{1.2808\times10^{-4} \pm \sqrt{4.6436\times10^{-7} \omega - 9.8014\times10^{-9}}}{2.3218\times10^{-7}} 
\end{equation}

\section{Discussion}
\label{discussion}
Our results and analysis are important steps in understanding grain size effects on carbonaceous chondrites over an expanded wavelength range. Some carbonaceous chondrite meteorites have been linked to C-type asteroids based on their low albedo and spectral similarities \citep{hiroi1993evidence,hamilton2019evidence,demeo2022connecting}. However, these linkages are conducted with carbonaceous chondrite powders, effectively assuming the surfaces of these asteroids are covered with fine-grained regolith like Ceres. This may be adequate for larger main-belt asteroids, though recent spacecraft missions to small, carbonaceous NEAs have depicted surfaces nearly devoid of regolith and riddled with boulders \citep{lauretta2019unexpected,watanabe2019hayabusa2}. Multiple experiments have shown that differences in grain size can affect spectral parameters \citep{reddy2015mineralogy,kiddell2018spectral}, including reflectance, slope, and band parameters. As a result, comparing the coarse surface of NEAs to meteorite powders may be unsuitable and lead to less accurate surface understanding. Given this potential problem, the central aim of our study was to analyze the effects of grain size on carbonaceous chondrite spectra to better address carbonaceous NEA interpretation. \\

VNIR reflectance spectroscopy with ground-based telescopes is a useful method of characterizing NEAs, as the atmosphere is largely transparent and a significant amount of information can be derived from absorption bands in this wavelength region, especially for silicate-dominated asteroids. Within this range, our results agree with \citet{kiddell2018spectral} and \citet{cloutis2018spectral}: increasing the grain size of carbonaceous chondrites generally results in a shallower spectral slope and decreased reflectance. With this, the diagnostic $\sim$1.0 $\mu$m olivine/pyroxene absorption feature remains present throughout all grain sizes for most of our meteorites. The 1.0 $\mu$m band depth of CM2 Aguas Zarcas, however, decreases and approaches the detection limit ($<$0.5\% depth) with grain sizes above 300 $\mu$m. This result has interesting implications for the classification of course-grained NEAs. The best link between C-type asteroids and carbonaceous chondrites is between CM chondrites and Ch-asteroids due to a distinct 0.7 $\mu$m hydration feature evident in both \citep{vilas1989phyllosilicate,hiroi1993evidence,burbine1998could,carvano2003search,fornasier2014aqueous,takir2013nature}. More recently, there has also been evidence of CR1 chondrites exhibiting this 0.7 $\mu$m feature \citep{beck2018controlling}. While this feature isn't always observed on hydrated asteroids, it is possible that a pre-existing 0.7 $\mu$m feature on a hydrated asteroid can disappear when thermally metamorphosed \citep{cloutis2012metamorphism}. Given just the VNIR spectrum of an object, the lack of diagnostic 0.7, 1.0, and 2.0 $\mu$m absorption features may result in ambiguous classifications, despite the object being compositionally-identical to material typically classified as a Ch-type. In this case, the classification may rely more on the spectral slope, radar observations, or evidence of hydration further into the infrared such as the $\sim$3.0 $\mu$m OH/H$_2$O band or additional MIR features discussed below. Recently, \citet{eschrig2021spectral} found a negative correlation between spectral slope at 2.0 $\mu$m and metamorphic grade, though this trend has only been shown in more silicate-dominated type 3 carbonaceous chondrites. \\

The 3.0 $\mu$m band is one of the more defining spectral features observed in carbonaceous chondrites that is a result of hydrated phyllosilicates within the matrix. Unlike the 0.7 $\mu$m hydration feature, the lack of a 3.0 $\mu$m absorption can be used to confirm the lack of hydration on the surface of an object (e.g. \citet{takir2020near}). Along with feature identification, multiple studies have found a correlation between degree of aqueous alteration and the depth of the 3.0 $\mu$m band \citep{beck2014transmission,king2015characterising,takir2015toward,garenne2016bidirectional,takir2019hydration,potin2020style,eschrig2021spectral}. While its importance in the characterization of carbonaceous chondrites is evident, the nature of the band makes it more difficult to measure in laboratory settings due to the effects of absorbed water. With this, the Earth’s atmosphere is largely opaque in this region due to atmospheric water, making ground-based observations especially difficult for fainter objects. While our samples all display prominent 3.0 $\mu$m features, we can’t comment on the effect of grain size since our measurements were conducted in ambient conditions. \\

MIR features past $\sim$8 $\mu$m are valuable in that they are unaffected by absorbed water and fall in the Earth's atmospheric transparency window reaching out to 14 $\mu$m \citep{salisbury1992emissivity}. As a result, we were able to study the effects of grain size on the two prominent features in this region: the Christiansen feature (C.F.) and Si-O stretching band (Si-O S.B.). The shift of the C.F. to shorter wavelengths with increasing grain size aligns with previous results \citep{hunt1972variation,logan1973compositional,henderson1997near,shirley2019particle}, though we specifically confirm this trend for carbonaceous chondrite-like material. The C.F. wavelength shift of up to -0.65 $\mu$m with increasing grain sizes may pose issues in the interpretation of hydration on remote bodies, as the C.F. location has been linked to aqueous alteration in carbonaceous chondrite powders and slabs \citep{mcadam2015aqueous,bates2020linking,hanna2020distinguishing}. Like the C.F., the position of the Si-O S.B. has been used to estimate the degree of aqueous alteration. \citet{mcadam2015aqueous} and \citet{bates2020linking} found that the 10-13 $\mu$m region in carbonaceous chondrite powders may be used to constrain alteration due to a combination of phyllosilicate and olivine vibrations. With less aqueous alteration, the longer-wavelength olivine feature near $\sim$12.3 $\mu$m is the most pronounced, while more extensive alteration results in the shift to Mg-rich phyllosilicates centered closer to $\sim$11.4 $\mu$m. \citet{hanna2020distinguishing} noticed a similar trend with polished carbonaceous chondrite slabs, but at slightly shorter wavelengths that better match our data (9.8-11.4 $\mu$m). Since \citet{mcadam2015aqueous} and \citet{bates2020linking} used powders that were $\leq$35 $\mu$m, we suspect that the location of this feature can be affected by transparency features only present in very fine powders with volume scattering \citep{shirley2019particle,hanna2020distinguishing}. Unlike the C.F. center, we did not observe a clear trend in the Si-O S.B. peak center with increasing grain size. Additionally, we see less variation in the Si-O S.B. center across grain sizes, with a wavelength shift of only -0.25 $\mu$m at larger grain sizes and slabs. Most of our meteorites only show a wavelength shift of $\pm$0.05 $\mu$m across our grain sizes, making the location of this feature possibly a more resistant metric of hydration across grain size than the C.F. center. Ground-based telescopes typically do not characterize NEAs in this higher wavelength range, though the BepiColombo spacecraft and JWST both have infrared capabilities for planetary surfaces. Currently, the OTES and OVIRS spectrometers on OSIRIS-REx can acquire spectra from 0.4-4.3 $\mu$m and 5.71-100 $\mu$m \citep{christensen2018osiris,reuter2018osiris}. \\

Much of the taxonomic classification of NEAs today relies on principal component analysis outlined by \citet{demeo2009extension}. As listed in the Analysis section, the PC values for our meteorites across all grain sizes are plotted in Figure \ref{fig:PCA}. Most notably, all grain size variation for each meteorite is generally parallel to the $\alpha$-line that separates typically feature-rich S-type spectra from typically featureless C/X-type spectra. This implies that grain variations on carbonaceous asteroids cannot move the asteroid’s PC values across the $\alpha$-line and result in an erroneous classification, similar to the space weather trends. The most extensively studied form of space weathering is lunar-style space weathering – an alteration of the surface of airless bodies owing to prolonged exposure to the space environment of our solar system (e.g., \citet{gaffey2010space}). Lunar-style space weathering affects meteorite and asteroid spectra by lowering their albedo, suppressing diagnostic VNIR bands, and reddening the spectral slope. This style of space weathering is understood to be caused by micrometeorite impacts, solar wind implantation, and other space environment factors forming nanophase iron particles on the surface of the asteroid regolith (e.g., \citet{pieters2016space}). \\

Space weathering on carbonaceous asteroid surfaces is not as well understood and several studies, both telescopic surveys and laboratory experiments, have provided conflicting results about the trends that space weathering on carbonaceous bodies will produce. Some evidence points toward carbonaceous bodies reddening and darkening \citep{hiroi2013keys,gillis2017incremental}, similar to lunar-style space weathering, while other evidence suggests carbonaceous material will weather to become increasingly bluer and darker over time \citep{moroz2004optical,matsuoka2015pulse,thompson2020effect}. \citet{trang2021role} found that the difference could be what nanophase particle manifests on the body’s surface as a result of the weathering. They found that carbonaceous chondrites that reddened with laboratory space weathering created nanophase iron, similar to lunar-style space weathering. Carbonaceous chondrites that became bluer with space weathering produced nanophase magnetite instead \citep{trang2021role}. \\

\citet{lantz2018space} measured principal component values for several carbonaceous chondrites before and after they were artificially weathered in the lab as described in \citet{lantz2017ion}. They found that these carbonaceous samples weathered differently depending on their composition. Anhydrous silicate materials tended to redden while organic-rich and hydrous materials in the meteorite became bluer \citep{lantz2018space}. Although the trend varied depending on the composition, the space weathering observed still altered the meteorites parallel to the alpha line in PCA space. It is important to understand the grain size effects on the carbonaceous meteorite PC values to better understand carbonaceous asteroid surfaces. The fact that the grain size trend overlaps with the space weathering trend of \citet{lantz2018space} implies that these two effects will be difficult to disentangle. \\

The variation of Bus-DeMeo asteroid taxonomic classification across grain size (Table \ref{table:asteroidtaxonomy}) also presents unique applications to NEAs, such as B-type asteroid Bennu \citep{hergenrother2013lightcurve}. We find that CM2 Aguas Zarcas, an appropriate spectral analog to asteroid Bennu \citep{hamilton2019evidence}, shifts from a Ch to a B-type classification at larger ($>$300 $\mu$m) grain sizes. \citet{ruggiu2021visible} found a similar result, with multiple carbonaceous chondrite raw slabs being matched to B-type asteroids due to their bluer slopes. The coarse surface of Bennu, seen from the OSIRIS-REx spacecraft \citep{lauretta2019unexpected}, may explain Bennu's classification as a B-type asteroid. This suggests that smaller carbonaceous asteroids with larger grain sizes, including low-albedo NEAs, may be taxonomically classified as B-types than compositionally identical and larger C-type asteroids in the main belt with smaller grain sizes/finer regolith.

\section{Summary}
We studied the UV-MIR (0.2-14 $\mu$m) reflectance spectroscopy of seven carbonaceous chondrite meteorites constrained to five different grain sizes (45-90 $\mu$m, 90-150 $\mu$m, 150-300 $\mu$m, 300-500 $\mu$m, and 500-1000 $\mu$m). We first plotted our results in terms of relative and absolute reflectance. We then analyzed how grain size affected absolute reflectance, spectral slope, and band parameters. From here, we used PCA analysis to find trends in PC space and see how grain size might affect asteroid classification with the Bus-DeMeo classification scheme. In addition, we performed Hapke modeling to investigate the effects of grain size on single scattering albedo for one of our meteorites, Aguas Zarcas. From our results and analysis, we find:

\begin{enumerate}
    \item Increasing grain size results in a decreased overall absolute reflectance
    \item Increasing grain size results in a decreased (bluer) spectral slope
    \item Increasing grain size affects 1.0 $\mu$m band depth for certain meteorite types with a significant weakening of the 1.0 $\mu$m band depth in CM2 Aguas Zarcas
    \item Increasing grain size results in a shift in the C.F. position to shorter wavelengths for most of our samples
    \item Increasing grain size generally results in less variation in the Si-O stretching band position than the C.F., possibly making it a more useful gauge of hydration if grain size is not constrained
    \item Like lunar-style space weathering, grain size effects in the principal component space are parallel to the $\alpha$-line
    \item Increasing grain size affects Bus-DeMeo asteroid taxonomic classification for C2-UNG Tarda and CM2 Aguas Zarcas, specifically shifting Aguas Zarcas from a Ch-type to a B-type
    \item With Hapke modeling, we present equations for constraining the grain size of Aguas Zarcas using single scattering albedo
\end{enumerate}

Our results may aid in the interpretation of small, C-type NEAs, including the actively studied asteroids Bennu and Ryugu. The near-absence of the 1.0 $\mu$m band depth in CM2 Aguas Zarcas at larger grain sizes means that Ch-type asteroids may have more ambiguous classifications, especially if thermally metamorphosed. While grain size can have a significant effect on the Christiansen feature position in carbonaceous chondrites, the Si-O stretching band center shows less variation across grain size, possibly making it more applicable in constraining aqueous alteration of remote bodies. Because grain size effects are parallel to the $\alpha$-line in PC space, it is unlikely that grain size itself can significantly affect the Bus-DeMeo classification of carbonaceous asteroids. The parallel nature of grain size variation to space weathering in PC space may result in difficulty separating the two effects in an object’s spectrum. 

\section*{Acknowledgments}
This research was funded by NASA Solar System Observations grant 80NSSC20K0632 (PI: Reddy) and NASA Near-Earth Object Observation program grant NNX17AJ19G (PI: Reddy). Data acquisition and processing were done in Tucson, AZ which is on the land and territories of Indigenous Peoples. We acknowledge our presence on the ancestral lands of the Tohono O'odham Nation and the Pascua Yaqui Tribe who have stewarded this area since time immemorial. Taxonomic type results presented in this work were determined, in whole or in part, using a Bus-DeMeo Taxonomy Classification Web tool by Stephen M. Slivan, developed at MIT with the support of National Science Foundation Grant 0506716 and NASA Grant NAG5-12355. Sections of this work were completed as part of the PTYS 520 Meteorites class (Spring 2022) taught by Reddy at the Lunar and Planetary Laboratory, University of Arizona, and is presented as an extension of the work of \citet{bowen2023grain} that focuses on ordinary chondrite and HED meteorites.

% \footnotesize{
%\bibliographystyle{plainnat}
%\bibliography{main}

\bibliographystyle{aasjournal}

\end{document}